\documentclass[prb,twocolumn,showpacs]{revtex4}

\AtBeginDvi{}

\usepackage{mathptmx}
\usepackage{amsmath,latexsym}
\usepackage{graphicx}
\usepackage[colorlinks,citecolor=blue]{hyperref}

\renewcommand*{\k}{\vec{k}}
\newcommand*{\kk}{\vec{k'}}
\newcommand*{\ks}{\vec{k}\sigma}
\newcommand*{\e}{\varepsilon}
\newcommand*{\ek}{\varepsilon_{\vec{k}}}
\newcommand*{\w}{\omega}

\newcommand*{\abs}[1]{\lvert #1\rvert}
\newcommand*{\expect}[1]{\langle #1\rangle}
\newcommand*{\correlation}[2]{\langle\!\langle #1 \vert #2 \rangle\!\rangle}
\newcommand*{\etal}{\emph{et al.}}

\DeclareMathOperator{\re}{Re}
\DeclareMathOperator{\im}{Im}

\providecommand{\texorpdfstring}[2]{#1}

\begin{document}

\title{Influence of disorder on the transport properties of heavy-fermion systems}

\author{Claas Grenzebach}
\author{Frithjof B.~Anders}
\author{Gerd Czycholl}
\affiliation{Institut f\"ur Theoretische Physik, Universit\"at Bremen,
             P.O. Box 330 440, D-28334 Bremen, Germany}

\author{Thomas Pruschke}
\affiliation{Institut f\"ur Theoretische Physik,
             Universit\"at G\"ottingen, D-37077 G\"ottingen, Germany}

\begin{abstract}
The influence of substitutional disorder on the transport properties of
heavy-fermion systems is investigated. We extend the dynamical mean-field
theory treatment of the periodic Anderson model (PAM) to a coherent-potential
approximation for disordered strongly correlated electron systems.
Considering two distinct local environments of a binary alloy $A_{c}B_{1-c}$
with arbitrary concentration $c$, we explore two types of disorder: on the
$f$ site and on the ligand sites. We calculate the spectral functions and
self-energies for the disordered PAM as well as the temperature dependence
of the resistivity and the thermoelectric power. The characteristic
concentration dependence as well as the order of magnitude of transport
properties are reproduced for metallic heavy-fermion systems and Kondo
insulators. In particular, sign changes of the Seebeck coefficient as
function of temperature and concentration are observed.
\end{abstract}

\maketitle

\section{Introduction}

Heavy-fermion systems (HFSs) are characterized by quasiparticles with a
very large effective mass at low temperatures $T$. This behavior occurs
in many lanthanide (rare-earth) and actinide compounds and has its origin
in the hybridization between the conduction bands and the local moments
of incompletely filled $f$ shells of the lanthanide or actinide
ions.\cite{Stewart01,Grewe91} These $f$ electrons contribute substantially
to the formation of the heavy quasiparticles. Measurements of the transport
coefficients on HFSs reveal characteristic anomalies at low temperatures.
The temperature dependence of the resistivity $\rho(T)$ of \emph{metallic} HFSs
(such as CePd$_3$, CeAl$_3$, CeCu$_6$, CeCu$_2$Si$_2$, UPd$_2$Al$_3$, etc.; see
Refs.~\onlinecite{Scoboria79,andres.graebner.ott.1975,ott.rudigier.etal.1984,
Onuki87,Grewe91,degiorgi.1999}) shows a rapid increase with increasing $T$
starting from a small residual value, which can often be fitted by a $T^2$
law. The resistivity saturates at $T_\textrm{max}$ and decreases with
increasing $T$ for $T>T_\textrm{max}$ as $\rho(T)\sim\log(T_\textrm{max}/T)$,
characteristic of the Kondo effect. In \emph{Kondo
insulators},\cite{degiorgi.1999,fisk.sarrao.1995,fisk.sarrao.1996} however, a
narrow gap opens\cite{logan.2003} at the Fermi energy at low temperatures. In
such materials, e.g., SmB$_6$ or Ce$_3$Bi$_4$Pt$_3$, the resistivity shows an
activation behavior for $T \to 0$. Another interesting transport quantity is
the thermoelectric power or Seebeck coefficient $S(T)$. At temperatures
comparable to $T_\textrm{max}$, the thermoelectric power can reach absolut
values of about $50\;\mu\mathrm{V/K}$, often accompanied with sign changes at
intermediate temperatures.\cite{Stewart01,Grewe91} In Kondo insulators, even
larger values of the thermoelectric power up to $300\;\mu\mathrm{V/K}$ have
been observed,\cite{jones.regan.disalvo.1998,jones.regan.disalvo.1999}
which may be interesting for low-temperature thermoelectric cooling.

In this paper, we examine alloys of heavy-fermion materials with arbitrary
concentrations. Disorder is introduced by substitution of ligand ions
(\emph{disorder on the ligand sites}) or of lanthanide or actinide ions
(\emph{disorder on the $f$ site}). A special case of disorder on the
$f$ site is given by replacing $f$-electron ions (e.g., Ce) by a certain
amount of nonmagnetic impurities (e.g., La) referred to as ``Kondo holes.''
While we focus on local disorder only, band disorder in the Kondo lattice
model has been recently considered\cite{BurdinFulde2007} using a
slave-boson mean-field approach.

Disorder has strong impact on the transport properties such as $\rho(T)$ and
$S(T)$. In substitutional alloys such as
La$_{1-x}$Ce$_x$Pd$_3$,\cite{Scoboria79}
Ce$_x$La$_{1-x}$Cu$_6$,\cite{Onuki87}
Ce$_x$La$_{1-x}$Cu$_{2.05}$Si$_2$,\cite{ocko.buschinger.etal.1999,ocko.drobac.etal.2001}
U$_{1-x}$Th$_x$Pd$_2$Al$_3$,\cite{plessis99}, etc.,
the residual resistivity $\rho(0)$ rapidly increases with increasing
concentration $x$ of the nonmagnetic impurities, and one obtains a crossover
from the metallic $\rho(T)$ behavior with a maximum to a monotonic curve,
where $\rho(T)$ decreases with increasing $T$. The thermoelectric power
$S(T)$ for CeCu$_{2.05}$Si$_2$ shows a crossover from one case with a
negative minimum, a sign change, and a positive maximum to a behavior with
two positive maxima and no sign change when substituting Ce by
La.\cite{ocko.buschinger.etal.1999,ocko.drobac.etal.2001}
For materials such as Ce$_3$Cu$_x$Pt$_{3-x}$Sb$_4$, ligand alloying
introduces a transition from a Kondo insulator to a dirty
metal.\cite{jones.regan.disalvo.1999}

The basic model for a description of the electronic properties of HFSs is the
periodic Anderson model\cite{varma.yafet.1976} (PAM). In a recent
paper,\cite{pam.2006} we studied the PAM within the dynamical mean-field
theory\cite{Pruschke95,Georges96} (DMFT) and showed that the characteristic
transport properties of HFSs can be understood within this framework. Within
the DMFT, the lattice model is mapped on an effective single-impurity Anderson
model\cite{Anderson61} (SIAM) by a self-consistency condition. We used the
numerical renormalization group\cite{Wilson75,krishna-murthy.wilkins.wilson.1980}
(NRG) as impurity solver for this effective SIAM. The NRG is a nonperturbative
method applicable at very low temperatures which reproduces the correct
characteristic low-temperature scale (Kondo temperature).\cite{PruschkeBullaJarrell2000}

In the present paper, we extend the DMFT-NRG treatment to the disordered
PAM. The disorder is studied within the coherent-potential
approximation\cite{elliott.krumhansl.leath.1974,yonezawa.1982} (CPA).
The CPA was originally developed for alloys modeled by noninteracting
particles. In the CPA, lattice coherence is restored by introducing an
average potential. Jani{\v s} and Vollhardt have pointed out that the
local nature of the potential allows one to embed the CPA into the DMFT
framework\cite{janis.vollhardt.1992} emphasizing that the CPA is the
best possible single-site approximation\cite{yonezawa.1982} for disorder.
On the other hand, the DMFT can be interpreted as CPA for periodic correlated
electron systems as the DMFT reduces to the CPA equations in the absence of
electron-electron interactions.\cite{vlaming.vollhardt.1992} This allows
for generalizations of the CPA to disordered systems with finite Coulomb
repulsion which has been exploited in the context of the Hubbard
model.\cite{janis.vollhardt.1992,janis.ulmke.vollhardt.1993,
ulmke.janis.vollhardt.1995,denteneer.ulmke.scalettar.ua.1998,
byczuk.hofstetter.vollhardt.2004}

There exists a number of previous works in which the influence of disorder
on the electronic properties of HFSs has been studied.\cite{leder.czycholl.1979,
CzychollLeder81,yoshimori.kasai.1986,xu.li.1990,freytag.keller.1990,li.qiu.1991,
Schlottmann92,wermbter.sabel.czycholl.1996,mutou.2001a,mutou.2001b}
Most of these papers consider the case of disorder by Kondo holes only.
The earliest work\cite{leder.czycholl.1979,CzychollLeder81} applied the
alloy analog approximation to the PAM as a CPA application.
Other treatments\cite{xu.li.1990,freytag.keller.1990,li.qiu.1991} used the
slave-boson mean-field approximation to the PAM, which enabled a
straightforward application of the standard
CPA.\cite{elliott.krumhansl.leath.1974,yonezawa.1982}
Schlottmann\cite{Schlottmann92} studied the development of impurity bands
within the hybridization gap of Kondo insulators in the low-concentration
limit using the iterative pertubation theory (ITP).
Mutou\cite{mutou.2001a,mutou.2001b} and Wermbter
\etal\cite{wermbter.sabel.czycholl.1996} calculated transport properties
within the framework of the ITP and second-order perturbation theory,
respectively.
Recently, the low-temperature behavior of paramagnetic Kondo lattices upon
random depletion of the local $f$ moments has been investigated by Kaul and
Vojta.\cite{KaulVojta2007}

We extend the DMFT in order to describe a disordered PAM, based on the
CPA for noninteracting systems in the spirit of Jani{\v s} and
Vollhardt.\cite{janis.vollhardt.1992} In contrast to the standard CPA
with static potentials, we use dynamical local potentials which include
correlation effects.
For reasons of simplicity, we call this extension ``CPA'', too, and stick to
the name ``DMFT'' for pure systems without disorder.
Here, we apply such a generalized CPA to the disordered PAM and calculate the
transport quantities $\rho(T)$ and $S(T)$ for disorder on the $f$ site as well
as for disorder on the ligand sites.
We treat the whole range of impurity concentrations $c\in[0;1]$ for different
choices of the PAM parameters, starting with either a metallic HFS or a Kondo
insulator for $c=0$. We find that this CPA-NRG treatment of the PAM is able
to explain the strong disorder dependence of the transport quantities of HFSs.

A generalization of the CPA for binary alloys to infinitely many local
environments has been considered in the context of ligand disorder of
UCu$_{5-x}$Pd$_x$ by Dobrosavljevi{\'c} and co-workers.\cite{PhysRevLett.69.1113,
miranda.dobrosavljevic.kotliar.1996,miranda.dobrosavljevic.kotliar.1997}
Those locally different environments yield a distribution of hybridization
strengths between the actinide and the conduction band at each site and, as
a consequence, a distribution of Kondo scales. By averaging over these
different low-energy scales, a non-Fermi-liquid behavior of transport
properties was obtained.\cite{miranda.dobrosavljevic.kotliar.1996,
miranda.dobrosavljevic.kotliar.1997,chattopadhyay.jarrell.1997}

HFSs often exhibit symmetry broken phases such as
superconductivity\cite{steglich.etal.1979} or
antiferromagnetism,\cite{Grewe91} which are certainly not yet properly
understood. Such phase transitions at finite temperatures can be
qualitatively understood within the PAM,\cite{GreweWelslau88,Grewe88,
WelslauGrewe1992,1DKondo1997,JarrellPangCox96,
tahvildar-zadeh.jarrell.freericks.1997,
doradzinski.spalek.1997,doradzinski.spalek.1998,Anders99,Anders2002}
but the complete phase diagram of the model in three dimensions has not yet
been determined. A vanishing transition temperature as a function of an
external control parameter such as magnetic field, pressure, or doping defines
the quantum critical point (QCP), at which the transition is governed by
quantum rather than thermal fluctuations. This sparked a lot of
experimental\cite{Stewart01} as well as 
theoretical\cite{Hertz76,Millis93,Sachdev2001} interest since it is believed
that these fluctuations strongly affect the transport properties at a finite
temperature in a cone close to the QCP.\cite{Millis93,Sachdev2001}
A proper microscopic description of such QCP requires the treatment of
one- and two-particle properties on equal footing, which is beyond the scope
of the present investigation and must be left for future research. Therefore,
we restrict ourselves to the influence of disorder on transport properties
in the paramagnetic phase of the model.

The paper is organized as follows: In Sec.~\ref{sec:theory}, we introduce the
model and the notations as well as our calculation methods, the DMFT, the CPA,
and the NRG as impurity solver.
Section~\ref{sec:Kondoholes} is devoted to the special case of Kondo holes:
in Sec.~\ref{sec:Kondoholes-AnalyticalConsiderations}, simplifications of the
CPA equations are considered, and in Sec.~\ref{sec:Kondoholes-SpectralProperties},
we discuss the temperature-dependent spectral functions and self-energy of the
PAM. These single-particle properties determine the transport properties via
the Kubo formulas introduced in Sec.~\ref{sec:transport}. We explicitly state
equations for the resistivity $\rho$ and the thermoelectric power $S$. In
Sec.~\ref{sec:transport-results}, results on the temperature dependence of
$\rho(T)$ and $S(T)$ are reported, both for disorder on the $f$ site and
disorder on the ligand sites. We conclude with a summary and outlook in
Sec.~\ref{sec:conclusion}.

\section{Theory}
\label{sec:theory}

The essential features of heavy-fermion compounds are based on the interplay
of localized, strongly correlated $f$ electrons with broad conduction bands.
At high temperature, the weakly coupled $f$ electrons cause mainly incoherent
and, with decreasing temperature, logarithmically growing spin-flip scattering
for the conduction electrons. Below a characteristic temperature scale, a
crossover to a coherent low-temperature phase is observed. The $f$ electrons
contribute significantly to the formation of heavy quasiparticles, while their
moments are dynamically screened. The PAM takes these ingredients into account,
comprising of spin degenerate conduction electrons, a lattice of correlated
localized $f$ electrons, and a hybridization (cf.\ Sec.~\ref{sec:model}).

In this work, we investigate the influence of disorder in HFSs. In general,
alloying with different ingredients $A$, $B$, $C$, \ldots, with concentrations
$c_A$, $c_B$, $c_C$, \ldots, respectively, destroys the lattice periodicity.
Here, we consider only substitutional alloys of the type $A_cB_{1-c}$ with
concentrations $c_A=c$ and $c_B=1-c$.

\subsection{Model}
\label{sec:model}

This section is divided into two parts: In the first part, we will introduce
the model for periodic systems, which is extended to binary alloys in the
second part.

\subsubsection{No disorder.}
The Hamiltonian of the simplest version of the PAM is given
by\cite{varma.yafet.1976}
\begin{align}
\label{eq:pam-hamil}
  \hat{H} &= \sum_{\ks} \e_{\ks} c^\dagger_{\ks} c_{\ks}
    + \frac{U}{2}\sum_{i\sigma} \hat{n}^f_{i\sigma} \hat{n}^f_{i-\sigma} \\
  &\quad+ \sum_{i\sigma}
    \begin{pmatrix}
      f^\dagger_{i\sigma} & c^\dagger_{i\sigma}
    \end{pmatrix}
    \mathcal{V}_{\sigma}
    \begin{pmatrix}
      f_{i\sigma} \\ c_{i\sigma}
    \end{pmatrix}
  \textrm{ with }\mathcal{V}_{\sigma} =
    \begin{pmatrix}
      \e_{f\sigma} & V_\sigma \\ V_\sigma & \e_{c\sigma}
    \end{pmatrix}.
\notag
\end{align}
Here,
$c_{\ks}$ ($c^\dagger_{\ks}$) destroys (creates) a conduction electron
with spin $\sigma$, momentum $\k$, and energy $\e_{\ks}+\e_{c\sigma}$,
where $\e_{c\sigma}$ indicates the band center.
The energy $\e_{f\sigma}$ denotes the spin-dependent single-particle
$f$-level energy at lattice site $i$,
$\hat{n}^f_{i\sigma} = f^\dagger_{i\sigma}f_{i\sigma}$ is the $f$-electron
occupation operator (per site and spin),
$f_{i\sigma}$ ($f^\dagger_{i\sigma}$) destroys (creates) an $f$ electron
with spin $\sigma$ at site $i$,
and $U$ denotes the on-site Coulomb repulsion between two $f$ electrons on
the same site $i$. The uncorrelated conduction electrons
hybridize locally with the $f$ electrons via the matrix element $V_\sigma$.

Even though only a single effective $f$ level is considered, this
model is quite general. It describes any heavy-fermion system
with odd ground-state filling of the $f$ shell, for which in a strong
crystal field environment the degenerate Hund's rule ground state
may be reduced to an effective spin-degenerate Kramers doublet. In
addition, charge fluctuations to even $f$ fillings leave the $f$ shell
in crystal field singlets.
The Hamiltonian contains four energy scales. The interplay between
$\e_{f\sigma}$ and $U$ controls the average $f$ filling as well as the local
moment formation for large $U$ and negative $\e_{f\sigma}$. The Anderson
width $\Gamma_0=V^2\pi\rho_0(0)$ determines the charge fluctuation scale of
the $f$ electrons, with $\rho_0(0)$ being the density of states of the
noninteracting conduction band of width $D$ at its band center.

The total filling per site, $n_\textrm{tot} =
\sum_{\sigma}(\expect{\hat{n}^c_{i\sigma}} + \expect{\hat{n}^f_{i\sigma}})$,
is kept constant by a temperature-dependent chemical potential $\mu(T)$. We
absorb the energy shifts into the band center $\e_{c\sigma}$ of the conduction
band, as well as the $f$ level $\e_{f\sigma}$.
For $n_\textrm{tot} = 2$ and $U=0$, the uncorrelated system is an insulator at
$T=0$, since the lower of the two hybridized bands is completely filled.
According to Luttinger's theorem, a finite $U$ of arbitrary strength does not
change the Fermi volume, which includes the full first Brillouin zone. As long
as the ground state does not change the symmetry due to a phase transition, the
system remains an insulator at arbitrarily large Coulomb repulsion. Therefore,
the nonmetallic ground state of Kondo insulators is not correlation induced,
but it is already present for the noninteracting system and is a consequence
of Luttinger's theorem. For nonintegral values of $n_\textrm{tot}$, the
paramagnetic phase of the system must be metallic.

\subsubsection{Binary alloy.}
The simplest version (\ref{eq:pam-hamil}) of the PAM can be extended to
describe two subsystems by replacing the matrix $\mathcal{V}_{\sigma}$ by a
random potential $\mathcal{V}_{i\sigma}$. At each lattice site $i$, two
different values are possible with a probability $c$ that is given by the
relative concentration of subsystem $A$:
\begin{equation}
\label{eq:random-potential}
  \mathcal{V}_{i\sigma} =
  \begin{cases}
    \mathcal{V}^A_\sigma & \textrm{with probability }c \\
    \mathcal{V}^B_\sigma & \textrm{with probability }1{-}c\textrm{.}
  \end{cases}
\end{equation}

In the case of no doping (i.e., $\mathcal{V}^A_\sigma = \mathcal{V}^B_\sigma$
or $c=0$ or $c=1$), the extended model reduces to Eq.~(\ref{eq:pam-hamil}).

While Eq.~(\ref{eq:pam-hamil}) includes possible Zeeman splitting of
the energies in an external magnetic field $H$, we set $H=0$
throughout the remainder of the paper and treat all properties as
spin degenerate. In particular, we drop the spin index $\sigma$
from now on for the sake of simplicity.

\subsection{Dynamical mean-field theory}
\label{sec:dmft}

Setting aside exact solutions in one dimension\cite{1DKondo1997}
using the Bethe ansatz for the Kondo lattice model, to our knowledge no
exact analytical solution has been found for the model
(\ref{eq:pam-hamil}) with finite $U$. Therefore, one has to rely on
suitable approximations for the PAM.
An obvious first approximation is the assumption of a purely local,
site-diagonal (i.e., $\vec{k}$ independent) self-energy, which for the PAM is
even better justified than for other lattice models of correlated electron
systems, as the first corrections are at least of order $V^6$.
Within a local self-energy approximation, the complicated lattice model can be
mapped on an effective SIAM.\cite{Anderson61} Such a mapping was first used
about 20 years ago in connection with the applications of the
noncrossing approximation\cite{Grewe83,Kuramoto83,Bickers87,KeiterKimball71}
(NCA) to the PAM.\cite{Kuramoto85,Kim87,Kim90,Grewe87,GreweKeiterPruschke88}
The effective site is viewed as a correlated atomic problem within a
time-dependent (or energy-dependent) external
field\cite{Kuramoto85,Grewe87,GreweKeiterPruschke88,BrandtMielsch89}
which is determined self-consistently, thus accounting for the feedback of the
electron propagation through the lattice.\cite{Kuramoto85,Kim87,Kim90}
Using the scaling of the tight-binding hopping parameter in large
dimensions by Metzner and Vollhardt,\cite{metzner.vollhardt.1989}
M\"uller-Hartmann\cite{MuellerHartmann89} has proven that the local self-energy
approximation becomes exact in the limit of infinite spatial dimensions.
Therefore, the self-consistency condition of the
DMFT,\cite{Kuramoto85,Jarrell92,GeorgesKotliar92,Pruschke95,Georges96}
which neglects spatial fluctuations in the single-particle self-energy,
becomes exact in the limit $d\to\infty$. Within weak-coupling $U$-perturbation
theory, it could be shown\cite{SchweitzerCzycholl90b,SchweitzerCzycholl91a}
that a local, $\vec{k}$-independent self-energy is a good approximation for
realistic dimension $d = 3$ as corrections due to intersite contributions to
the self-energy are negligibly small. As a consequence, phase transitions
remain mean-field-like in DMFT since $\k$-dependent fluctuations are not
included in a local approximation.\cite{Anders99,Georges96,Anders2002}

The following exact relations for the conduction electron Green
function $G_c(\k,z)$ and the $f$-electron Green function
$G_f(\k,z)$ can be obtained for the PAM (\ref{eq:pam-hamil}):
\begin{align}
  G(\k,z) &=
  \begin{bmatrix}
    G_{ff}(\k,z) & G_{fc}(\k,z) \\ G_{cf}(\k,z) & G_{cc}(\k,z)
  \end{bmatrix}
\notag \\
  &= \left\{
  \begin{pmatrix}
    z & 0 \\ 0 & z-\ek
  \end{pmatrix}
  -
  \begin{bmatrix}
    \e_f+\Sigma_f(\k,z) & V \\ V & \e_c
  \end{bmatrix}
  \right\}^{-1},
\label{eq:conduction-electrons}
\end{align}
where $z$ is any complex energy off the real axis.
Within a local approximation such as the DMFT, the $\k$-dependent
$f$-electron self-energy $\Sigma_f(\k, z)$ is replaced by a
$\k$-independent $\Sigma_f(z)$.
From Eq.~(\ref{eq:conduction-electrons}), one defines a self-energy
of the conduction electrons via
\begin{equation}
\label{eq:self-energy-gc}
  \Sigma_c(z) = \frac{V^2}{ z -\e_f -\Sigma_f(z)},
\end{equation}
which can include a simple $\k$ dependence through the hybridization
matrix elements $V^2$, here taken as a constant.
For such a local self-energy $\Sigma_c$, the
site-diagonal conduction-electron Green function $G_{cc}$ can be
written as a Hilbert transformation
\begin{equation}
\label{eqn:g-loc}
  G_{cc}(z) = \frac{1}{N} \sum_{\k}G_{cc}(\k,z) = D[z-\e_c-\Sigma_c(z)],
\end{equation}
defined for arbitrary complex argument $z$ as
\begin{equation}
\label{eqn:hilbert-tranform}
  D(z) = \int_{-\infty}^{\infty} d\e \frac{\rho_0(\e)}{z-\e},
\end{equation}
where $\rho_0(\w)$ is the density of states of the noninteracting
conduction electrons.

The DMFT self-consistency condition states that the site-diagonal
matrix element of the $f$-electron Green function of the PAM must
be equal to $G^{\textrm{loc}}_{ff}(z)$ of an effective site problem
\begin{align}
\label{eq:pam-scc}
  G_{ff}(z) &= \frac{1}{N} \sum_{\k}G_{ff}(\k,z) = G^{\textrm{loc}}_{ff}(z), \\
  G^{\textrm{loc}}_{ff}(z) &=  \frac{1}{z-\e_f-\Delta(z)-\Sigma_f(z)},
\label{eq:f-gf-local}
\end{align}
with the same local $f$-electron self-energy $\Sigma_f(z)$
for the lattice and the effective site. This defines the
self-consistency condition for the functions $\Sigma_f(z)$
and $\Delta(z)$.

We can put the site-diagonal Green functions into a Green function matrix
\begin{subequations}
\label{eq:gf-matrix}
\begin{equation}
  G(z) =
  \begin{bmatrix}
    G_{ff}(z) & G_{fc}(z) \\ G_{cf}(z) & G_{cc}(z)
  \end{bmatrix},
\end{equation}
but it is sufficient to know one of the components because they are connected
to each other:
\begin{equation}
  G_{fc}(z) = G_{cf}(z)
  = \frac{\Sigma_c(z)}{V}G_{cc}(z)
  = \frac{\Delta(z)}{V}G_{ff}(z).
\end{equation}
\end{subequations}
Furthermore, the self-consistency condition (\ref{eq:pam-scc}) extends to the
full matrix, $G(z) = G^{\textrm{loc}}(z)$. It is possible to
formulate the DMFT equation for the conduction-electron Green function
instead of the $f$-electron Green function:
\begin{equation}
  \Delta(z) = z-\e_f-\Sigma_f(z)-G_{ff}(z)^{-1}
  = \frac{V^2}{\Sigma_c(z)+G_{cc}(z)^{-1}}.
\end{equation}
This condition can also be written in the following form:
\begin{equation}
  \begin{pmatrix}
    z & 0 \\ 0 & z
  \end{pmatrix}
  -\begin{bmatrix}
    \e_f+\Sigma_f(z) & V \\ V & \e_c
  \end{bmatrix}
  - G^{-1}(z) =
  \begin{bmatrix}
    0 & 0 \\ 0 & z-\e_c-\tfrac{V^2}{\Delta(z)}
  \end{bmatrix}.
\label{eq:dmft}
\end{equation}

Given the Green functions $G_{ff}(z)$ and $G_{cc}(z)$, their spectral functions
\begin{subequations}
\begin{align}
  \rho_f(\w) &= \im G_{ff}(\w-i 0^+) / \pi, \\
  \rho_c(\w) &= \im G_{cc}(\w-i 0^+) / \pi
\end{align}
\label{eq:specfuncDMFT}
\end{subequations}
determine the local occupation numbers
\begin{equation}
  n_{f/c} = \sum_\sigma \int_{-\infty}^\infty d\w \; f(\w-\mu) \rho_{f/c}(\w),
\label{eq:occDMFT}
\end{equation}
where the spin sum contributes only a factor of $2$ in the absence of a
magnetic field and $f(\w)$ denotes the Fermi function. Then, the total filling
per site is given by $n_\textrm{tot} = n_f +n_c$. Particle-hole symmetry is
reached at $n_\textrm{tot}=2$ and $\e_f-\e_c=-U/2$ for a symmetric $\rho_0(\w)$.
As a matter of convenience, we will perform an integral transformation such
that $\mu$ is absorbed into $\e_f$ and the band center $\e_c$; all energies
will be measured with respect to $\mu$. For a given lattice filling
$n_\textrm{tot}$, we have to adjust $\mu$ in addition to fulfill
Eq.~(\ref{eq:pam-scc}).

Before we discuss the solution of the effective site, let us briefly comment
on the implications of the analytical form of the conduction-electron
self-energy (\ref{eq:self-energy-gc}).
For a Fermi liquid at $T\to 0$, the imaginary part of $\Sigma_f$
vanishes quadratically close to the chemical potential, i.e.,
$\im\Sigma_f(\w-i0^+) \propto \w^2$.
At particle-hole symmetry, $\Sigma_c(z)$ diverges as $1/z$, leading to an
insulator. Away from particle-hole symmetry, the denominator remains
finite and the imaginary part also must have Fermi-liquid properties
$\im\Sigma_c(\w-i0^+) \propto \im\Sigma_f(\w-i0^+) \propto \w^2$.
The real part is  very large, and therefore, the spectral function
$\rho_c(\w)$ as well as $\rho_f(\w)$ sample the high-energy band edges of
$\rho_0(\w)$, yielding a hybridization gap.
These analytic properties must be fulfilled by any approximate
solution of the DMFT self-consistency condition (\ref{eq:pam-scc})
consistent with Fermi-liquid theory.

\subsection{Coherent-potential approximation}
\label{sec:cpa}

The DMFT is only applicable to periodic systems. When we consider alloys or
doped systems where the lattice periodicity is destroyed, then the DMFT has
to be extended. There are several possible approximations\cite{yonezawa.1982}
for disordered alloys, e.g., the virtual-crystal method, the average t-matrix
approximation,\cite{CoxGrewe88} or the CPA.\cite{elliott.krumhansl.leath.1974,
yonezawa.1982,janis.vollhardt.2001,byczuk.hofstetter.vollhardt.2004} The CPA is
the best single-site approximation\cite{yonezawa.1982,vlaming.vollhardt.1992}
and can be seen as an extension of the DMFT for two subsystems. If there is
effectively only one subsystem (as in the case of $\mathcal{V}_A=\mathcal{V}_B$
or $c=0$ or $c=1$), the CPA reduces to the DMFT.

The standard CPA self-consistency condition for noncorrelated systems is given by
\begin{multline}
  c \left\{ \openone - [\mathcal{V}_A-\Sigma^\textrm{CPA}]\mathcal{G} \right\}^{-1}
    [\mathcal{V}_A-\Sigma^\textrm{CPA}] \\
  + (1-c) \left\{ \openone - [\mathcal{V}_B-\Sigma^\textrm{CPA}]\mathcal{G} \right\}^{-1}
    [\mathcal{V}_B-\Sigma^\textrm{CPA}]
  = 0,
\label{eq:standardCPA}
\end{multline}
where $\Sigma^\textrm{CPA}(z)$ is the self-energy (matrix) of the
configurationally averaged Green function (matrix)
\begin{align}
  \mathcal{G}(z) &=
  \begin{bmatrix}
    \mathcal{G}_{ff}(z) & \mathcal{G}_{fc}(z) \\
    \mathcal{G}_{cf}(z) & \mathcal{G}_{cc}(z)
  \end{bmatrix}
\notag \\
  &= \frac{1}{N}\sum\limits_{\vec{k}}\left[
  \begin{pmatrix}
    z & 0 \\ 0 & z-\ek
  \end{pmatrix}
  - \Sigma^\textrm{CPA}(z) \right]^{-1}
\label{eq:configuration-average}
\end{align}
[as in Eq.~(\ref{eqn:g-loc}), the CPA Green function matrix can also be written
as a Hilbert transformation by replacing the $\k$ sum with the integral $D$].

To include correlation effects in Eq.~(\ref{eq:standardCPA}), the potential
matrix $\mathcal{V}_{A/B}$ of subsystems $A$ and $B$, respectively, has to
be replaced by the sum
\begin{equation}
  \mathcal{V}_{A/B} \leadsto \mathcal{V}_{A/B} +
  \begin{bmatrix}
    \Sigma^{A/B}_f(z) & 0 \\
    0 & 0
  \end{bmatrix},
\end{equation}
where $\Sigma^{A/B}_f(z)$ is the local $f$-electron self-energy of subsystems
$A$ and $B$, respectively.
Equivalent to Eq.~(\ref{eq:standardCPA}) with correlation is the following
CPA equation:
\begin{equation}
  \mathcal{G}(z) = c G^A(z) + (1-c) G^B(z),
\label{eq:cpa}
\end{equation}
where $G^A$ and $G^B$ are the local Green function matrices (\ref{eq:gf-matrix})
of effective site problems for the two subsystems $A$ and $B$,
given by the effective media $\Delta_A(z)$ and $\Delta_B(z)$, respectively:
\begin{subequations}
\label{eq:gf-a-b}
\begin{gather}
  G^{A/B}_{ff}(z) = \bigl[z-\e^{A/B}_f-\Delta_{A/B}(z)-\Sigma^{A/B}_f(z)\bigr]^{-1}, \\
  G^{A/B}_{fc}(z) = \frac{\Sigma^{A/B}_c(z)}{V_{A/B}}G^{A/B}_{cc}(z)
  = \frac{\Delta_{A/B}(z)}{V_{A/B}}G^{A/B}_{ff}(z).
\end{gather}
\end{subequations}

Note that the determinant of $\mathcal{G}$ [Eq.~(\ref{eq:configuration-average})]
can be written in the following simple form:
\begin{equation}
  \det \mathcal{G}(z) = \frac{\mathcal{G}_{cc}(z)}{z - \Sigma^\textrm{CPA}_{ff}}.
\end{equation}
Using this expression, it can be shown by direct calculation that
\begin{equation}
  \begin{pmatrix}
    z & 0 \\ 0 & z
  \end{pmatrix}
  - \Sigma^\textrm{CPA}(z) - \mathcal{G}^{-1}(z) =
  \begin{bmatrix}
    0 & 0 \\ 0 & \Gamma(z)
  \end{bmatrix}
\label{eq:gamma}
\end{equation}
has got one component only. Therefore, we obtain the same structure as in
the case of the DMFT, Eq.~(\ref{eq:dmft}). This leads to a DMFT-like
self-consistency for the effective medium by setting
\begin{equation}
  \Delta_{A/B}(z) := \frac{V^2_{A/B}}{z-\e^{A/B}_c-\Gamma(z)}.
\label{eq:medium-a-b}
\end{equation}
This ensures that for pure systems ($c=0$ or $c=1$) the DMFT limit as given by
Eq.~(\ref{eq:dmft}) is recovered.

As in Eqs.~(\ref{eq:specfuncDMFT}) and (\ref{eq:occDMFT}),
the CPA spectral functions
\begin{subequations}
\begin{align}
  \rho^\textrm{CPA}_f(\w) &= \im\mathcal{G}_{ff}(\w-i 0^+) / \pi, \\
  \rho^\textrm{CPA}_c(\w) &= \im\mathcal{G}_{cc}(\w-i 0^+) / \pi
\end{align}
\end{subequations}
determine the occupation numbers
\begin{equation}
  n_{f/c} = \sum_\sigma \int_{-\infty}^\infty d\w \; f(\w-\mu) \rho^\textrm{CPA}_{f/c}(\w).
\end{equation}
In analogy to Eq.~(\ref{eq:self-energy-gc}), we define
\begin{equation}
  \Sigma^\textrm{CPA}_c(z) :=
  \frac{\Sigma^\textrm{CPA}_{fc}(z)\Sigma^\textrm{CPA}_{cf}(z)}
    {z-\Sigma^\textrm{CPA}_{ff}(z)}.
\label{eq:self-energy-gc-CPA}
\end{equation}

The CPA self-consistency cycle consists of two parts:
(I) After solving the two impurity subsystems $A$ and $B$ for the given
effective media, the CPA self-energy matrix is determined by
Eqs.~(\ref{eq:cpa}) and (\ref{eq:gamma}).
(II) A new estimate for the CPA Green function matrix is obtained by
Eq.~(\ref{eq:configuration-average}). Both matrices are required to calculate
$\Gamma(z)$---and thus two new effective media---by Eq.~(\ref{eq:gamma}).
The cycle is completed by shifting $\e^{A/B}_f$ and $\e^{A/B}_c$ such that
$N[G]=n_\textrm{tot}$ and $\mu=0$.

Although the CPA includes the DMFT as end points for the concentrated
systems ($c=0$ and $c=1$), conceptual differences in disordered systems
arise from the occurrence of the additional quantities. In the CPA, we
must distinguish between the local matrix $\mathcal{V}_{A/B}$, which
describes the properties of a site of type $A/B$ and \emph{differs} at
different sites, and the configuration averaged $k$-independent lattice
self-energy $\Sigma^\textrm{CPA}(z)$, which is \emph{equal} for all lattice
sites. Moreover, the configurationally averaged CPA Green function
restores translational invariance in the lattice and is not equal to
either local Green function $G^{A/B}(z)$. Even though the CPA yields only
one configurationally averaged dynamical field $\Gamma(z)$, the local
dynamics of each site is determined by different media $\Delta_{A/B}(z)$.
Only for $c=0$ and $c=1$, the CPA equation (\ref{eq:cpa}) is trivially
fulfilled and $\Sigma^\textrm{CPA}(z)$ coincides with the local self-energy
as well as the CPA Green function $\mathcal{G}(z)$ coincides with the one of
the effective site.

\subsection{Impurity solver}
\label{sec:ImpSolver}

The local $f$-electron self-energy $\Sigma^{A/B}_f$ entering the CPA
equation (\ref{eq:cpa}) via the local Green function matrix $G^{A/B}$
[Eq.~(\ref{eq:gf-a-b})] is obtained for an effective site defined by the local
Hamiltonian\cite{Kim87,Jarrell92,GeorgesKotliar92,Pruschke95,Georges96,pam.2006}
$\hat{H}_\textrm{eff}^{A/B}$ for subsystems $A$ and $B$, respectively,
\begin{align}
\label{eq:effective-siam}
  \hat{H}_\textrm{eff}^{A/B}
  &= \sum_\sigma (\e^{A/B}_{f\sigma} -\mu) f^\dagger_\sigma f_\sigma
  + U \hat n_\uparrow \hat n_\downarrow
  + \sum_\sigma \int d\e\; (\e-\mu)
    d^\dagger_{\sigma\e}  d_{\sigma\e} \nonumber \\
  &\quad + \sum_\sigma \int d\e\; V_{A/B} \sqrt{\rho^{A/B}_\textrm{eff}(\e)}
    \left(
    d^\dagger_{\sigma\e} f_\sigma + f^\dagger_\sigma d_{\sigma\e}
    \right).
\end{align}
The coupling of the $f$ electron to a fictitious bath of
``conduction electrons'' created by $d^\dagger_{\sigma\e}$
is described by an energy-dependent hybridization function
\begin{subequations}
\begin{equation}
  \pi V^2_{A/B}\rho^{A/B}_\textrm{eff}(\e)
  = \im\Delta_{A/B}(\e-i0^+).
\end{equation}
Using Eq.~(\ref{eq:medium-a-b}), we can write the density of states
of the fictitious bath as
\begin{equation}
  \rho^{A/B}_\textrm{eff}(\e)
  = \frac{1}{\pi}\im[\e-i0^+ -\e^{A/B}_c-\Gamma(\e-i0^+)]^{-1}.
\end{equation}
\end{subequations}

We accurately solve the Hamiltonian (\ref{eq:effective-siam}) using Wilson's
NRG approach.\cite{Wilson75,krishna-murthy.wilkins.wilson.1980} The key
ingredient in the NRG is a logarithmic discretization of the continuous
bath, controlled by the parameter\cite{Wilson75} $\Lambda > 1$. The
Hamiltonian is mapped onto a semi-infinite chain, where the $N$th link
represents an exponentially decreasing energy scale $D_N\sim\Lambda^{-N/2}$.
Using this hierarchy of scales, the sequence of finite-size Hamiltonians
$\mathcal{H}_N$ for the $N$-site chain is solved iteratively, discarding the
high-energy states at each step to maintain a manageable number of
states. The reduced basis set of $\mathcal{H}_N$ thus obtained is
expected to faithfully describe the spectrum of the full Hamiltonian
on the scale of $D_N$, corresponding\cite{Wilson75} to a temperature
$T_N \sim D_N$ from which all thermodynamic expectation values are
calculated. The energy-dependent hybridization function $\Delta(z)$
determines the coefficients of the semi-infinite
chain.\cite{BullaPruschkeHewson97,pam.2006} For further details, we
refer to the recent NRG review by Bulla \etal\cite{BullaCostiPruschke2007}

The finite-temperature NRG spectral functions are calculated using the
recently developed algorithm from Refs.~\onlinecite{peters2006} and
\onlinecite{WeichselbaumDelft2007}.
The usage of a complete basis set of the Wilson chain originally derived
for real-time dynamics of quantum impurites out of
equilibrium\cite{AndersSchiller2005,AndersSchiller2006} ensures that
the spectral sum rule is exactly fulfilled and that the NRG occupancy is
accurately reproduced for arbitrary values of $\Lambda$ and number of
retained states $N_s$. The NRG spectrum is broadened in the usual
way\cite{BullaCostiVollhardt01,AndersCzycholl2004,BullaCostiPruschke2007}
by a Gaussian at higher frequency and a Lorentzian for low-frequency
excitations $\abs{E}$
\begin{subequations}
\label{eq:broadening}
\begin{equation}
  \delta(\w-E) \leadsto
  \begin{cases}
    \frac{e^{-b^2/4}}{\sqrt{\pi} b\abs{E}}
    e^{-[\log(\w/E)/b]^2}
      &\text{for } \abs{E}\ge L_w T \\
    \frac{1}{\pi} \frac{L_T T}{(\w-E)^2 + (L_T T)^2}
      &\text{for } \abs{E}<L_w T.
  \end{cases}
\end{equation}
As long as not stated otherwise, we use
\begin{align}
b&=0.6, &L_T&=1, &L_w &= \sqrt{\pi} L_T \exp(-b^2/4)/b.
\end{align}
\end{subequations}
The local $f$-electron self-energy $\Sigma^{A/B}_f$ is determined by the exact ratio
\begin{equation}
\label{eqn:sigma-nrg}
  \Sigma^f_\sigma(z) = U \frac{M^\textrm{NRG}_\sigma(z)}{F^\textrm{NRG}_\sigma(z)},
\end{equation}
derived via equation of motion technique,\cite{BullaHewsonPruschke98}
where the correlation functions $M^\textrm{NRG}_\sigma(z) =
\correlation{f_\sigma f^\dagger_{-\sigma} f_{-\sigma}}{f^\dagger_{\sigma}}(z)$
and $F^\textrm{NRG}_\sigma(z) =
\correlation{f_\sigma }{f^\dagger_{\sigma}}(z)$ have been obtained
from NRG spectral functions.

We use a Gaussian model density of states
\begin{equation}
  \rho_0(\e) = \frac{\exp[-(\e/t^*)^2/2]}{t^*\sqrt{2\pi}}
\label{eq:gauss}
\end{equation}
for the unperturbed conduction-electron system.
In the following, we set $\sqrt2 t^* = 10\Gamma_0$ and measure energies
in units of $\Gamma_0 = \pi V_0^2 \rho_0(0)$ which defines a reference
hybridization $V_0$ by $V_0^2= 2t^*\Gamma_0/\sqrt{2\pi} =
10\Gamma_0^2/\sqrt\pi \approx 5.64\Gamma_0^2$.
In order to make contact to the experimentally relevant parameter regime,
we assume $\Gamma_0=100\;\mathrm{meV}$, which translates into a
temperature scale of $\frac{\Gamma_0}{k_B}\approx 1160\;\mathrm{K}$.

\section{Kondo Holes}
\label{sec:Kondoholes}

\subsection{Analytical considerations}
\label{sec:Kondoholes-AnalyticalConsiderations}

A special case for disorder on the $f$ site is given by introducing
Kondo holes, i.e., lattice sites without $f$ electrons. We describe the
two subsystems with identical parameters
($V:=V_A=V_B$, $\e_c:=\e^A_c=\e^B_c$, and therefore $\Delta:=\Delta_A=\Delta_B$)
except that the $f$-level energy of system $B$ is shifted toward infinity.
Additionally, the total electron density is reduced by one electron per $B$ site
to $n_\textrm{tot}=n_\textrm{tot}(c{=}1)-(1-c)$.

Because there are no $f$ electrons in system $B$, the conduction-electron
self-energy $\Sigma_c^B$ [Eq.~(\ref{eq:self-energy-gc})] vanishes. By
Eqs.~(\ref{eq:gf-matrix}) and (\ref{eq:dmft}), we conclude that the
Green function matrix of system $B$ contains only the $cc$ element:
\begin{equation}
  G^B(z) =
  \begin{pmatrix}
    0 & 0 \\ 0 & \Delta(z)/V^2
  \end{pmatrix}.
\end{equation}
As a consequence, the configurationally averaged $f$-electron Green function
of Eq.~(\ref{eq:cpa}) is given by $\mathcal{G}_{ff}(z)=c G^A_{ff}(z)$,
vanishing for $c\to0$, and the configurationally averaged band-electron
Green function $\mathcal{G}_{cc}(z)$ interpolates between $G^A_{cc}(z)$ and
the free medium with band center at $\e_c$.

By inserting Eq.~(\ref{eq:cpa}) into Eq.~(\ref{eq:gamma}), we conclude that
in the case of Kondo holes the CPA self-energy matrix reduces to a scalar self-energy
\begin{subequations}
\label{eq:cpa-efB_infty-Sigmaf}
\begin{align}
  \Sigma^\textrm{CPA}(z) &=
  \begin{bmatrix}
    \Sigma^\textrm{CPA}_{ff}(z) & V \\ V & \e_c
  \end{bmatrix}
\intertext{ with}
  \Sigma^\textrm{CPA}_{ff}(z) &= z - [\mathcal{G}^{-1}(z)]_{ff}
    = z - \Delta_A(z) - \frac{1}{cG^A_{ff}(z)} \notag \\
    &= (1-\tfrac{1}{c}) [z - \Delta(z)] + \tfrac{1}{c} [e^A_f+\Sigma^A_f(z)].
\end{align}
\end{subequations}
There are two contributions to the imaginary part of $\Sigma^\textrm{CPA}_{ff}$:
a term $\frac{1}{c} \im\Sigma^A_f$ and a term $\frac{1-c}{c} \im\Delta$. If
$\Sigma^A_f$ remains Fermi-liquid-like for $\omega, T\to 0$, a finite life time
is introduced by the additional $\frac{1-c}{c} \im\Delta(0)$ for $c<1$.
We thus interpolate between the following DMFT limits:
\begin{subequations}
\begin{align}
  \lim_{c\to1} \left[c\Sigma^\textrm{CPA}_{ff}(z)\right]
  &= \varepsilon^A_f + \Sigma^A_f(z), \\
  \lim_{c\to0} \left[c\Sigma^\textrm{CPA}_{ff}(z)\right]
  &= -\frac{1}{G^A_{ff}(z)}, \\
  \lim_{c\to1} \left[\frac{1}{c}\Sigma^\textrm{CPA}_c(z)\right]
  &= \Sigma^A_c(z), \\
  \lim_{c\to0} \left[\frac{1}{c}\Sigma^\textrm{CPA}_c(z)\right]
  &= V^2 G^A_{ff}(z).
\end{align}
\end{subequations}

Because the CPA mixes two subsystems, it is not clear how to define
a low-temperature scale for the complete system. Each effective site has
its own low-temperature scale that characterizes the local dynamics.
In the case of Kondo holes, we are in a better situation. As stated
above, the configurationally averaged $f$-electron Green function
is directly connected to the $f$-electron Green function of subsystem $A$:
$\mathcal{G}_{ff}(z)=c G^A_{ff}(z)$, there are no $f$ electrons in
system $B$ and, hence, no special low-temperature effects.
Then we can take the low-temperature scale of subsystem $A$ to define a
low-temperature scale $T_\textrm{low}$ for the complete system
which is only defined up to a constant factor.
We will use the renormalization of the Anderson width by the
quasiparticle spectral weight
\begin{equation}
  \label{eq:qp-weight-t0}
  T_0 = \Gamma_0\left[1- \left. \frac{\partial
        \re\Sigma^A_f(\w)}{\partial \w}
    \right|_{\w,T\to 0}
  \right]^{-1}
\end{equation}
as our choice of such a low-temperature scale $T_0\propto T_\textrm{low}$
for our numerical analysis.\cite{vidhyadhiraja.logan.2004}
It is related to the mass enhancement $m^*/m = \Gamma_0/T_0$.

\subsection{Spectral properties}
\label{sec:Kondoholes-SpectralProperties}

In this section, we take a look at the spectral properties of a system with
Kondo holes.

\begin{figure}
  \includegraphics[width=0.8\columnwidth]{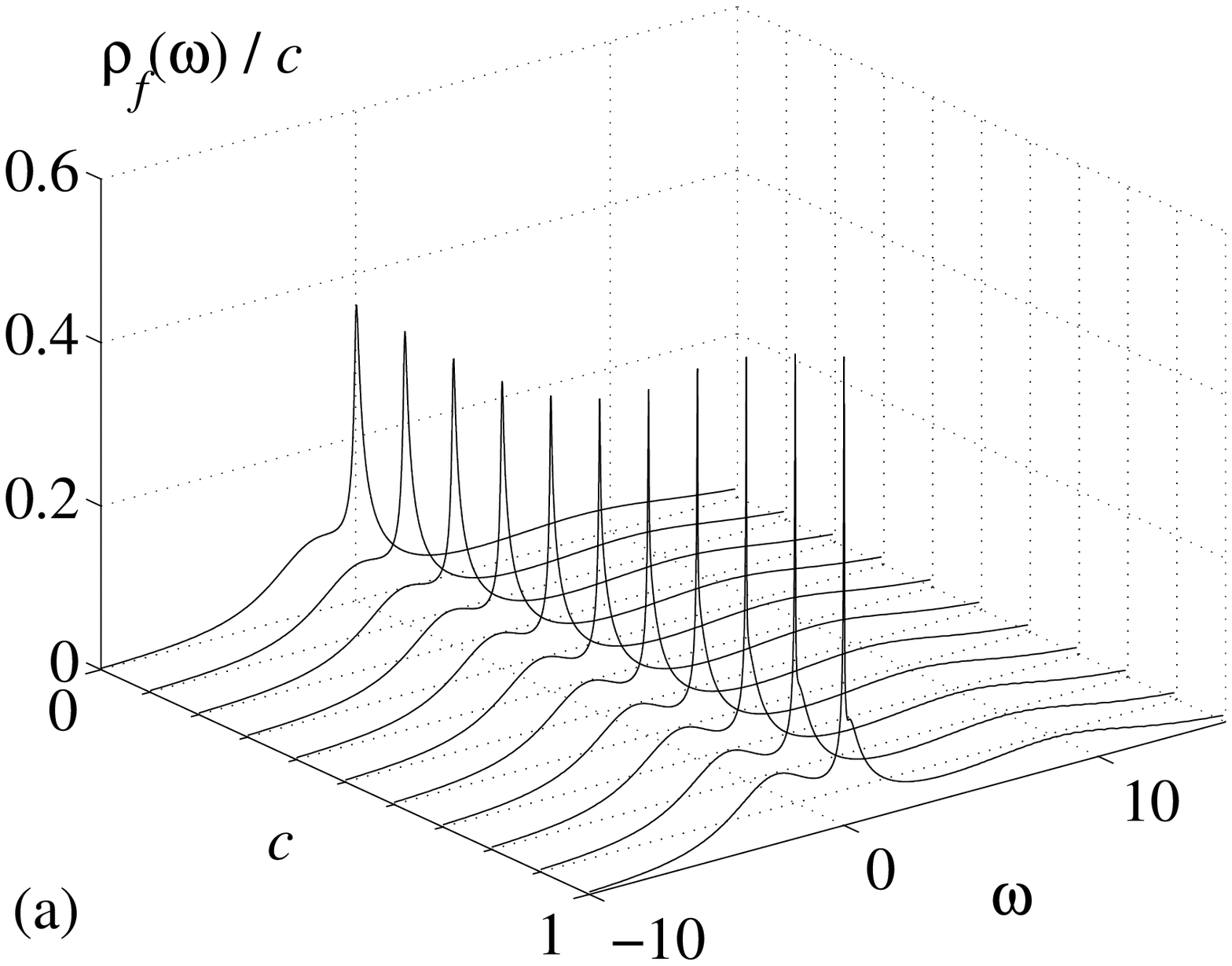}
  \includegraphics[width=0.8\columnwidth]{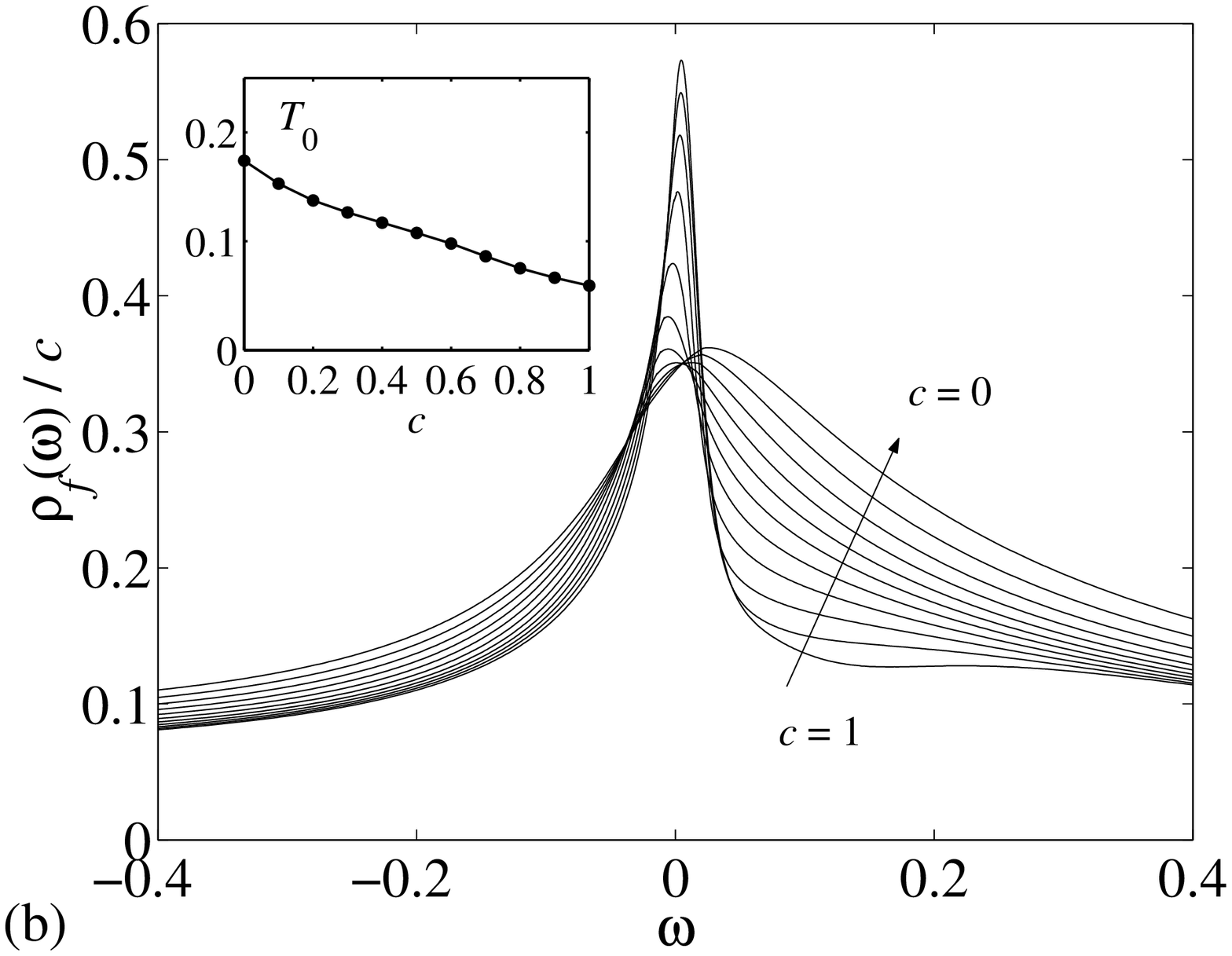}
  \caption{Spectral density $\rho_f(\w)$ for different concentrations
    of system $A$, $c\in\{10^{-4},0.1,0.2,0.3,0.4,0.5,0.6,0.7,0.8,0.9,1\}$,
    calculated with CPA-NRG for $V_A=V_B=V_0$, $U/\Gamma_0=10$,
    $\e^A_f-\e^A_c=-U/2$, $\e^B_f=\infty$, chemical potential $\mu=0$, and a
    filling $n_\textrm{tot}=1.6-(1-c)$ at a small temperature $0.2\;\mathrm{K}$.
    (a) Overview of the generic three-peak structure,
    (b) Peak at $\w=0$ in more detail with corresponding temperature scale
    $T_0$ [Eq.~(\ref{eq:qp-weight-t0})] in the inset.
    NRG parameters: number of retained NRG states, $N_s=800$,
    $\Lambda=1.6$, and $\delta/\Gamma_0=10^{-3}$.}
  \label{fig:ImGf-U10-smallT}
\end{figure}

The typical structure of the $f$-electron spectral function
$\frac{1}{c}\rho^\textrm{CPA}_f(\w)=\rho^A_f(\w)$
is shown in Fig.~\ref{fig:ImGf-U10-smallT}.
A pronounced peak structure dominates the low-energy part of the spectrum
in the vicinity of the chemical potential.
In addition, we observe two shallow high-energy peaks: one at $\e_f$
below $\mu$, and one at $\e_f + U$, which corresponds to double
occupancy of the $f$ levels. It is always correctly positioned by the
NRG, independent of the value of $U$, but with a linewidth too large
due to the NRG broadening procedure of Eq.~(\ref{eq:broadening})
(see Refs.~\onlinecite{AndersCzycholl2004} and
\onlinecite{BullaHewsonPruschke98} for details).

The dependence on the concentration $c$ is small, but the peak width is
enlarged with decreasing $c$. This indicates an increasing low-temperature
scale; indeed, the scale given by Eq.~(\ref{eq:qp-weight-t0}) and shown in the
inset of Fig.~\ref{fig:ImGf-U10-smallT}(b) increases with decreasing $c$.

\begin{figure}
  \includegraphics[width=0.8\columnwidth]{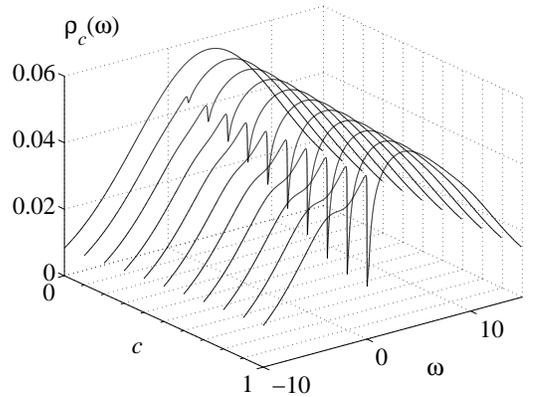}
  \caption{Spectral density $\rho_c(\w)$,
     with all parameters as in Fig.~\ref{fig:ImGf-U10-smallT}.
     With $c\to0$, the gap at $\w=0$ vanishes
     and we reach a Gaussian curve.}
  \label{fig:ImGc-U10-smallT}
\end{figure}

The band-electron spectral function is depicted in Fig.~\ref{fig:ImGc-U10-smallT}.
Equivalent to the vanishing of $f$ electrons from the system with decreasing $c$
is the disappearing of the gap structure and the crossover to a free medium of
Gaussian shape as discussed in the previous section.

\begin{figure}
  \includegraphics[width=0.8\columnwidth]{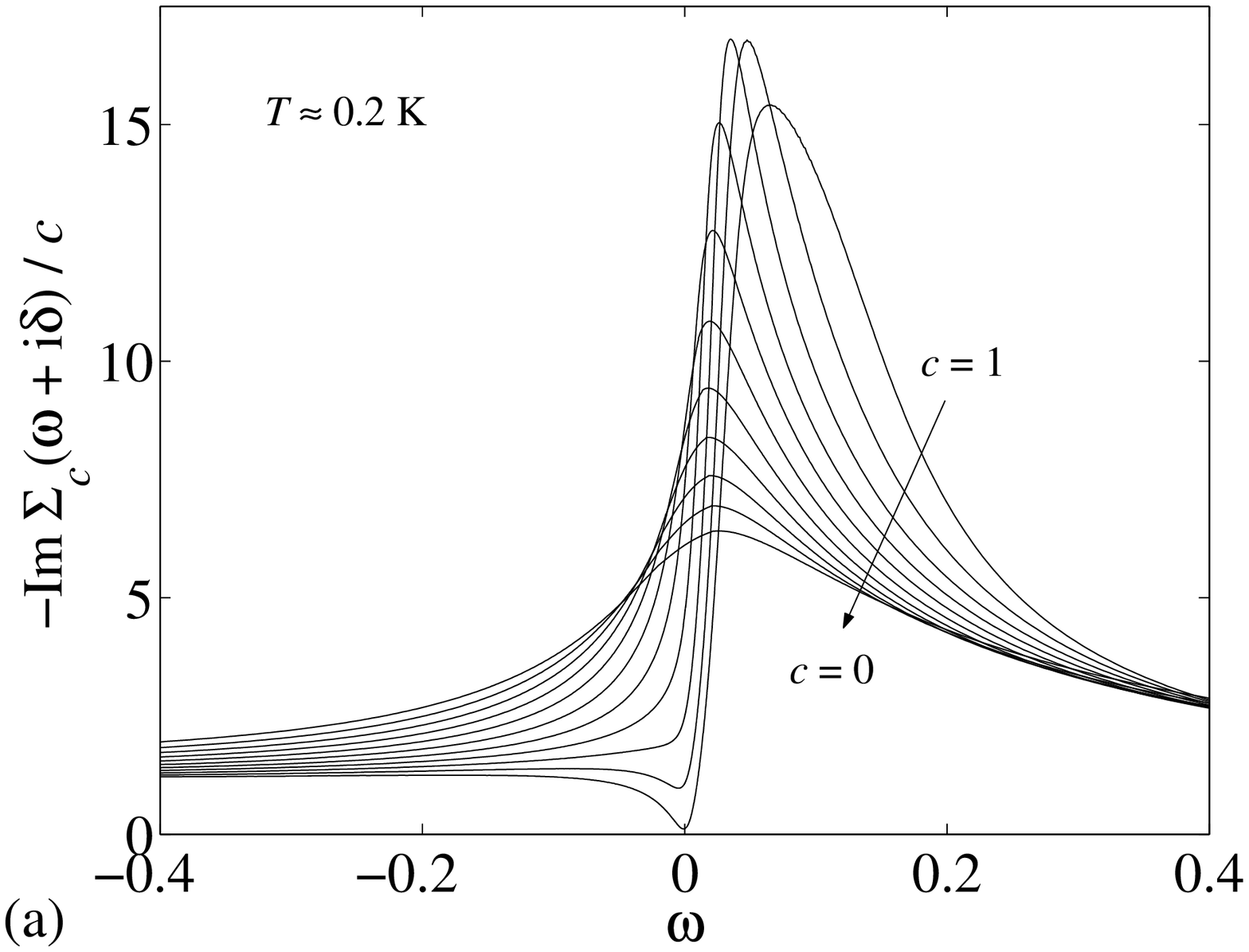}
  \includegraphics[width=0.8\columnwidth]{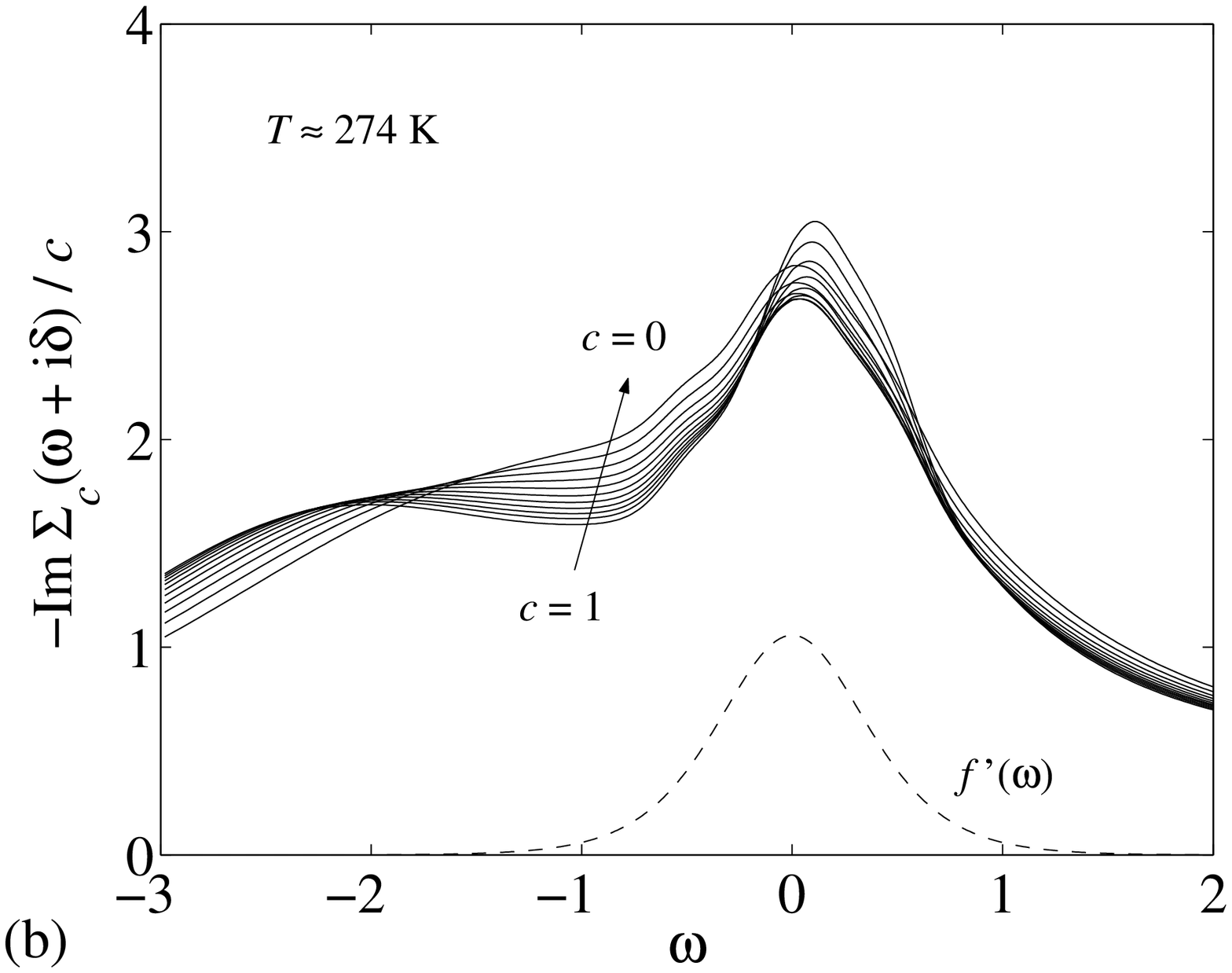}
  \caption{Band self-energy $\frac{1}{c}\Sigma^\textrm{CPA}_c$
    for different concentrations of system $A$,
    $c\in\{10^{-4},0.1,0.2,0.3,0.4,0.5,0.6,0.7,0.8,0.9,1\}$,
    and all parameters as in Fig.~\ref{fig:ImGf-U10-smallT}:
    (a) at $T\approx0.2\;\mathrm{K}$, (b) at $T\approx274\;\mathrm{K}$,
    with additional derivative of Fermi function, $f'$,
    showing the relevant part of the energy axis.}
  \label{fig:Sc_c-U10}
\end{figure}

The behavior of the spectral functions is reflected in the conduction-electron
self-energy $\frac{1}{c}\Sigma^\textrm{CPA}_c(z)$ (cf.\ Fig.~\ref{fig:Sc_c-U10}).
At low temperatures and $c=1$, we observe a characteristic $\w^2$ dependence at
the chemical potential $\mu=0$, indicating a Fermi liquid.
At $c=0$, the self-energy is proportional to $G^A_{ff}$.
The imaginary part of the conduction-electron self-energy at a temperature of
$274\;\mathrm{K}$ can be seen in Fig.~\ref{fig:Sc_c-U10}(b).
Additionally, the derivative of the Fermi function, $f'$, is shown. The product
of $f'$ and $\im \Sigma^\textrm{CPA}_c$ enters the transport integrals in
Sec.~\ref{sec:transport}. That is, only the part of $\im \Sigma^\textrm{CPA}_c$
around $\omega=0$ is relevant for transport calculations, and in this range,
the imaginary part of the conduction-electron self-energy is nearly independent
of the concentration $c$.

\section{Transport theory}
\label{sec:transport}

To describe the electronic transport within the PAM, we start from the
standard relations\cite{Mahan81} for the generalized transport coefficients,
according to which the electrical current density $\vec{J}$ and the heat
current density $\vec{q}$ depend linearly on the electric field $\vec{E}$
and the temperature gradient $\nabla T$:
\begin{subequations}
\begin{align}
  \vec{J} &= L_{11} \vec{E} + L_{12} \bigl(-\tfrac{1}{T}\nabla T\bigr), \\
  \vec{q} &= L_{21} \vec{E} + L_{22} \bigl(-\tfrac{1}{T}\nabla T\bigr).
\end{align}
\end{subequations}
All coefficients are calculated within the linear response approach,
starting from similar Kubo formulas.\cite{Luttinger64,Mahan81} For
symmetry reasons, $L_{12}=L_{21}$ must hold.

For example, the real part of the frequency-dependent (optical) conductivity
tensor\cite{Voruganti92,Mahan81,Czycholl2000} $\sigma(\w)=L_{11}(\w)$ is
related to the current-current correlation function and is written as
\begin{equation}
  \sigma_{\alpha\beta}(\w)
  =-\frac{1}{\w NV_0} \im\correlation{j_{\alpha}}{j_{\beta}^\dagger}(\w+i0^+),
\end{equation}
where $V_0 = a^3$ is the volume of the unit cell and $N$ counts the number
of lattice sites.
It has been shown\cite{CzychollLeder81} that the current operator of the PAM
has two contributions: a conduction-electron part and a part proportional to
$\nabla V_{\k}$. The (bare) $f$ electrons do not appear in the current, since
they do not disperse. For a $\k$-independent hybridization, only the conduction
electrons carry the electrical and heat currents as follows:
\begin{equation}
  \vec{j} = e\sum_{\ks} \vec{v}_{\k} c^\dagger_{\ks} c_{\ks},
\end{equation}
where $\vec{v}_{\k} = \frac{1}{\hbar}\nabla_{\k}\ek$ is the group velocity.
Hence, the current-susceptibility tensor $\correlation{\vec{j}}{\vec{j}^\dagger}(z)$
is connected to the particle-hole Green function
\begin{equation}
  \correlation{\vec{j}}{\vec{j}^\dagger}(z)
  = e^2\sum_{\sigma\sigma'\k\kk} \vec{v}_{\k} \vec{v}_{\kk}^T
  \correlation{c^\dagger_{\ks} c_{\ks}}{c^\dagger_{\kk\sigma'} c_{\kk\sigma'}}(z).
\end{equation}
In a cubic crystal, the conductivity is isotropic: $\sigma_{\alpha\beta}(\w)
= \sigma(\w)\openone$.
From now on, we will consider only the $xx$ component of the conductivity
$\sigma(\w) \equiv \sigma_{xx}(\w)$.

In general, the full two-particle Green function
$\correlation{c^\dagger_{\ks} c_{\ks}}{c^\dagger_{\kk\sigma'} c_{\kk\sigma'}}(z)$ 
involves  vertex corrections which reflect residual particle-particle
interactions.\cite{Mahan81} However, in the limit $d\to\infty$, it was shown that
current operator vertex corrections vanish.\cite{Khurana90,SchweitzerCzycholl91b}
Thus, it is consistent with the DMFT assumption of a $\k$-independent self-energy
that these vertex corrections vanish for any lattice model of correlated electron
systems. For the special case of a local approximation for the PAM, this was
already shown in Refs.~\onlinecite{LorekAndersGrewe91}, \onlinecite{AndersCox97},
and \onlinecite{CoxGrewe88}, as for symmetry reasons
$\sum_{\k} \vec{v}_{\k} \abs{V_{\k}}^2 G_{\k}(z+\w)G_{\k}(z) = 0$.
The CPA maintains the locality of the self-energy and we can neglect vertex corrections
in the case of the CPA, too (see also Ref.~\onlinecite{janis.vollhardt.2001}). All DMFT
formulas can be used for the CPA; we only have to substitute the DMFT self-energies
with CPA self-energies. In both cases, we obtain
\begin{align}
  \correlation{j_x}{j_x^\dagger}(\w+i0^+)
  &= \frac{2e^2}{\hbar^2} \sum_{\k}
  \left(\frac{\partial\ek}{\partial k_x} \right)^2
  \int_{-\infty}^\infty d\w' f(\w') \rho_c(\ek,\w') \nonumber \\
  &\quad\times[G_{cc}(\k,\w'+\w + i0^+) \nonumber \\
  &\quad+ G_{cc}(\k,\w'-\w - i0^+)].
\label{eq:-chi-jj-v}
\end{align}
Within the DMFT or the CPA, the lattice one-particle Green function depends
only on the (complex) energy $z$ and bare band dispersion $\ek$:
$G_{cc}(\k,z) = G_{cc}(\ek,z)$. Then
\begin{equation}
  \frac{1}{N}\sum_{\k} \left(\frac{\partial\ek}{\partial k_x} \right)^2
  A(\ek) = \int_{-\infty}^\infty d\e \,\tilde\rho_0(\e) A(\e),
\end{equation}
with
\begin{equation}
  \tilde\rho_0(\e) = \frac{1}{N}\sum_{\k}
  \left(\frac{\partial\ek}{\partial k_x} \right)^2
  \delta(\e -\ek).
\label{eq:rho-tilde}
\end{equation}
We use a Gaussian model density of states [Eq.~(\ref{eq:gauss})]
for the unperturbed conduction-electron system which is appropriate
for a $d$-dimensional hypercubic lattice in the limit
$d\to\infty$,\cite{metzner.vollhardt.1989} and
$\tilde\rho_0(\e)$ has been evaluated approximately in large
dimensions\cite{Pruschke93} as
\begin{equation}
  \tilde\rho_0(\e) = \frac{(at^*)^2}{d} \rho_0(\e) + O(d^{-2})
\end{equation}
on a hypercubic lattice. Then, Eq.~(\ref{eq:-chi-jj-v}) can be reduced to a
sum of Hilbert transforms, which is defined in Eq.~(\ref{eqn:hilbert-tranform}).

By taking the limit $\w\to 0$, the static conductivity $\sigma=L_{11}$
and the thermoelectric power
\begin{equation}
  S = \frac{1}{T}\frac{L_{12}}{L_{11}}
    = \frac{k_B}{e} \frac{eL_{12}}{k_B T L_{11}}
\label{eq:lin-res-thermo}
\end{equation}
are obtained.\cite{Czycholl2000,Mahan81}
The thermoelectric power $S$ is defined as the proportionality constant
between an applied temperature gradient and the measured voltage drop
in the absence of a current flow. The Peltier coefficient, given by the
ratio of heat and electrical current, is related to the thermoelectric
power by $\Pi=T S$.

In the limit $\w\to 0$, we get for the generalized transport coefficients
the following:
\begin{subequations}
\begin{align}
  (\sigma=)\, L_{11} &= \frac{1}{\hbar a}
    \int_{-\infty}^\infty[-f'(\w)]\ e^2\  \tau(\w)\ d\w,
\label{eq:cond} \\
  L_{12} &= \frac{1}{\hbar a}
    \int_{-\infty}^\infty[-f'(\w)]\ e\w\  \tau(\w)\ d\w,
\label{eq:L12} \\
  L_{22} &= \frac{1}{\hbar a}
    \int_{-\infty}^\infty[-f'(\w)]\ \w^2\ \tau(\w)\ d\w.
\end{align}
Here, $f'$ is the derivative of the Fermi function; thus, we have the limits
\begin{align}
  L_{11}(T{=}0) &= \frac{e^2}{\hbar a} \tau(0),
  &L_{12}(T{=}0) &= 0,
  &L_{22}(T{=}0) &= 0,
\label{eq:L_T0}
\end{align}
\end{subequations}
and $\tau(\w)$ represents a generalized relaxation time defined as
\begin{align}
  \tau(\w) &= \frac{2\pi}{d}(t^*)^2\int_{-\infty}^\infty \rho_0(\e) \rho^2_c(\e,\w) d\e
\notag \\
  &= \frac{(t^*)^2}{\pi d}
    \left[ \frac{\partial\im D(z)}{\partial\im z}
    - \frac{\im D(z)}{\im z}\right],
\label{equ:relaxation-time}
\end{align}
which is to be evaluated at $z=x+iy$
\begin{equation*}
  z = \begin{cases}
    \w+i0^+-\e_c-\Sigma_c(\w+i0^+)
      &\textrm{(DMFT)} \\
    \w+i0^+-\Sigma^\textrm{CPA}_{cc}(\w+i0^+) - \Sigma^\textrm{CPA}_c(\w+i0^+)
      &\textrm{(CPA).}
  \end{cases}
\end{equation*}
Here, $\e_c$ corresponds to the component $\Sigma^\textrm{CPA}_{cc}$,
and $\Sigma_c$ and $\Sigma^\textrm{CPA}_c$ are defined in
Eqs.~(\ref{eq:self-energy-gc}) and (\ref{eq:self-energy-gc-CPA}), respectively.
For small $y=\im z$, e.g., in the Fermi-liquid regime, the approximation
\begin{equation}
  \tau(\w) = \frac{(t^*)^2}{d} \left[\frac{\rho_0(x)}{y}
  + \frac{\rho_0^{\prime\prime}(x)}{2}y
  + O(y^3) \right]
\label{eq:relaxation-time-approx}
\end{equation}
is valid. In lowest order, we have $\tau(\w)\propto 1/\im\Sigma_c(\w+i0^+)$
and the linearized Boltzmann transport theory is recovered.
Though we are using the full form of Eq.~(\ref{equ:relaxation-time}) for calculations,
it is sufficient to consider the lowest order only for discussion of the
properties of the transport coefficients because it is the main contribution.

\section{Results for the transport properties}
\label{sec:transport-results}

All transport calculations rely on the results for the single-particle Green
functions of the PAM. As stated in Sec.~\ref{sec:ImpSolver}, we use a Gaussian
model density of states [Eq.~(\ref{eq:gauss}) with $\sqrt2 t^* = 10\Gamma_0$]
for the unperturbed conduction-electron system and measure energies in units
of $\Gamma_0=100\;\mathrm{meV}$. In particular, this defines a reference
hybridization $V_0$ by $V_0^2= 2t^*\Gamma_0/\sqrt{2\pi} =
10\Gamma_0^2/\sqrt\pi \approx 5.64\Gamma_0^2$.

If we assume one electron per unit cell of the volume $a^3$
($a=10^{-10}\;\mathrm{m}$), the resistivity $\rho=\sigma^{-1}$ has the
natural unit
\begin{equation}
  \sigma_0^{-1} = \hbar a/e^2 \approx 41\;\mu\Omega\;\mathrm{cm}.
\label{eq:sigma_0}
\end{equation}
Note that $eL_{12}/k_B T L_{11}$ is dimensionless and
$k_B/e\approx -86\;\mu\mathrm{V/K}$. Therefore, the thermoelectric
power is given in absolute units; only the scale of the temperature axis
must be fixed by experiment. With our choice of $\Gamma_0$, it is given in
units of $\frac{\Gamma_0}{k_B}\approx 1160\;\mathrm{K}$.

\subsection{Disorder on the \texorpdfstring{$f$}{f} site}

In this section, we restrict ourselves to the special case of Kondo holes,
as introduced in Sec.~\ref{sec:Kondoholes}. In particular,
the $f$-level energy of system $B$ is shifted to infinity.
The hybridization $V=V_A=V_B$ is set to the reference value $V_0$.

\subsubsection{Resistivity}

\begin{figure}
  \includegraphics[width=0.8\columnwidth]{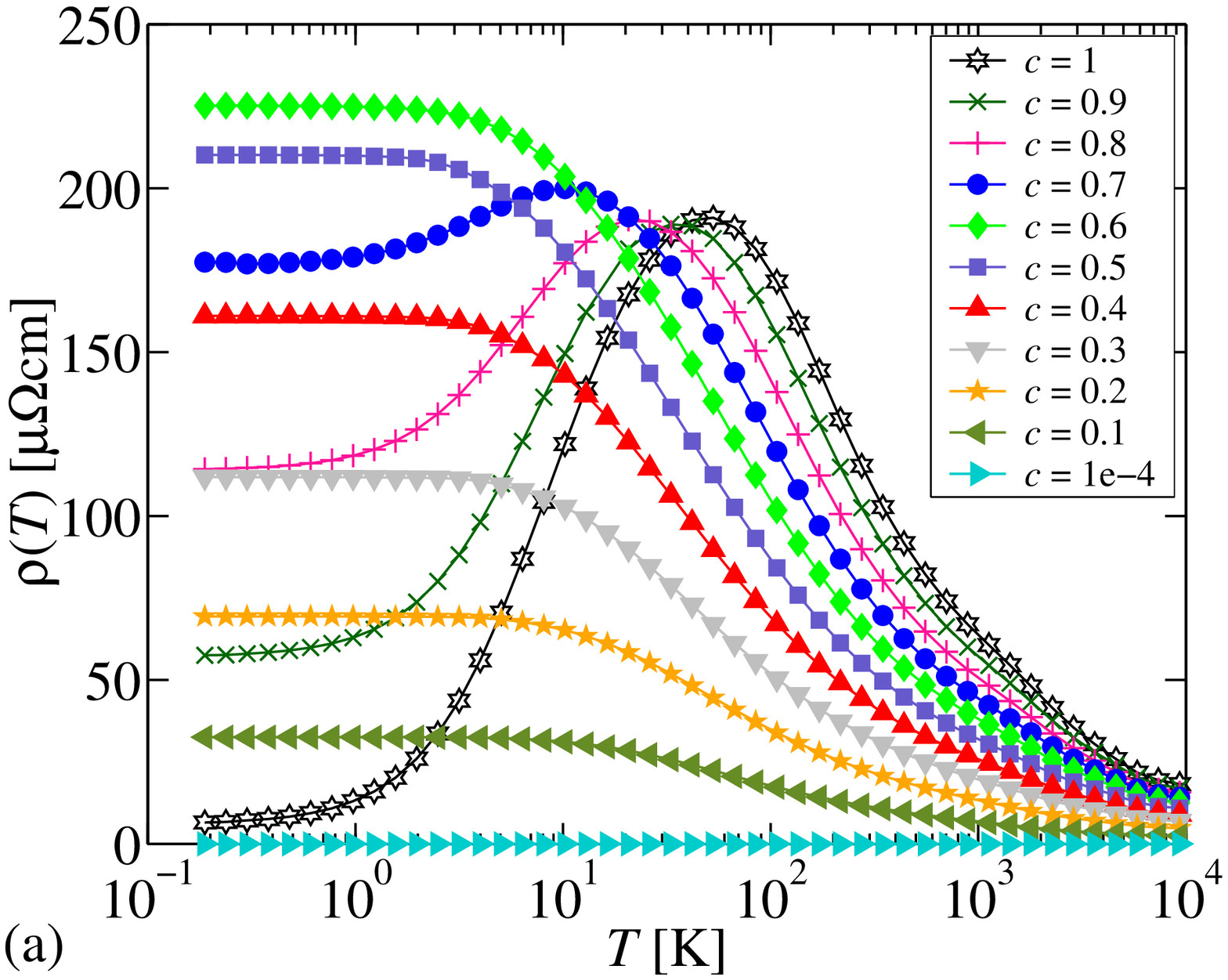}
  \includegraphics[width=0.8\columnwidth]{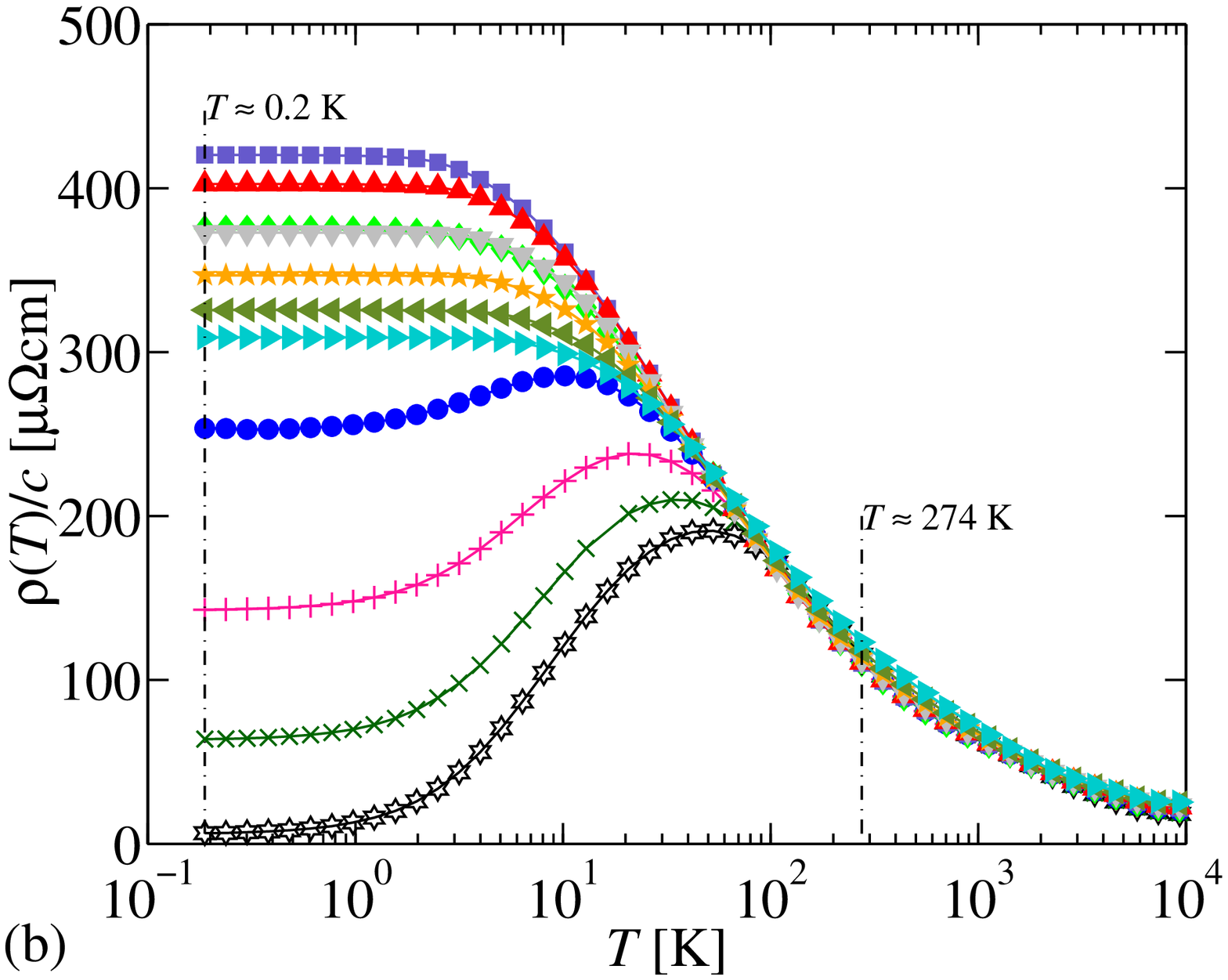}
  \caption{(Color online)
    Resistivity $\rho(T)$ as a function of $T$ for different
    concentrations $c$ of system $A$, calculated with CPA-NRG;
    all parameters as in Fig.~\ref{fig:ImGf-U10-smallT},
    i.e., $U/\Gamma_0=10$, $\e^A_f-\e^A_c=-U/2$, $\e^B_f=\infty$,
    and $n_\textrm{tot}=1.6-(1-c)$.
    In (b), the same data are plotted as $\rho(T)/c$.
    The dash-dotted lines mark two cuts through the resistivity curves
    which are shown in Fig.~\ref{fig:ftau-U10} in more detail.}
  \label{fig:rho-t-U10}
\end{figure}

In Fig.~\ref{fig:rho-t-U10}, the resistivity is displayed for
$U/\Gamma_0=10$, $\e^A_f-\e^A_c=-U/2$, and various values of the
concentration $c$ of system $A$.
The filling is set to $n_\textrm{tot}=1.6-(1-c)$.

In the pure case, $c=1$, we reproduce the typical behavior of metallic
heavy-fermion systems
\cite{Scoboria79, andres.graebner.ott.1975, ott.rudigier.etal.1984, Onuki87}
within our CPA-NRG treatment:
a resistivity increasing with increasing $T$ for low $T$, a maximum of the
order of $100\;\mu\Omega\;\mathrm{cm}$ at a characteristic temperature
$T_\textrm{max}$, and a $\rho(T)$ (logarithmically) decreasing with increasing
$T$ for $T > T_\textrm{max}$. Note that even for $c=1$, a finite $\rho(T=0)$ is
obtained for $T\to 0$. This is due to the fact that we are using a finite
$\delta=10^{-3}\Gamma_0$ contributing to the self-energy imaginary part. In
addition, the broadening of Eq.~(\ref{eq:broadening}) and the limited accuracy
of the NRG in the regime $\w < T$ further enhance the self-energy imaginary
part; for a discussion, see also Ref.~\onlinecite{pam.2006}.

\begin{figure}
  \includegraphics[width=0.9\columnwidth]{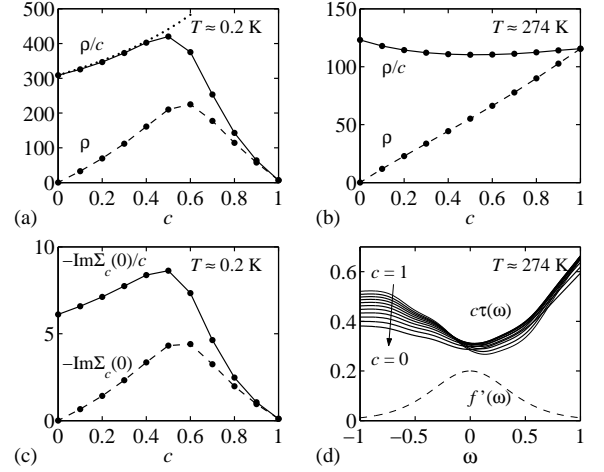}
  \caption{[(a) and (b)] Resistivity as a function of the concentration $c$
    at $T\approx 0.2\;\mathrm{K}$ and $T\approx 274\;\mathrm{K}$, respectively:
    $\rho(c)$ (dashed line), $\rho(c)/c$ (full line),
    and approximation $\propto 1/(1-0.6c)$ (dotted line).
    (c) Illustration of the proportionality
    $\rho(T\to0)/c \propto \im\Sigma^\textrm{CPA}_c(0)/c$
    [cf.\ Fig.~\ref{fig:Sc_c-U10}].
    (d) Generalized relaxation time $c\tau(\w)$ for $T\approx 274\;\mathrm{K}$.
    The additional Fermi derivation shows the relevant part of the energy axis.
    The temperatures correspond to dash-dotted lines in Fig.~\ref{fig:rho-t-U10}.}
\label{fig:ftau-U10}
\end{figure}

In the disordered case, $c<1$, we distinguish the two regimes
$T<T_\textrm{max}(c{=}1)$ and $T>T_\textrm{max}(c{=}1)$ to obtain a
qualitative classification for all concentrations $c$. At high temperatures
$T>T_\textrm{max}$, the logarithmic gradient of $\rho(T)$ has its origin in
the weak incoherent scattering of conduction electrons by the $f$ moments.
With increasing concentration $1-c$ of Kondo holes, there are less scattering
events. Indeed, the resistivity decreases and scales linearly with the
concentration $c$ as depicted in Fig.~\ref{fig:rho-t-U10}(b). The $c$
dependence of the resistivity $\rho$ as well as of $\rho/c$ at a fixed
temperature $T\approx274\;\mathrm{K}>T_\textrm{max}$ is plotted in greater
detail in Fig.~\ref{fig:ftau-U10}(b).

This can be understood using the transport equation
(\ref{eq:cond}) with abbreviation (\ref{eq:sigma_0}) as follows:
\begin{equation*}
  \frac{\rho}{c} = \sigma_0^{-1}
  \left(\int_{-\infty}^\infty[-f'(\w)] c\tau(\w) d\w \right)^{-1}.
\end{equation*}
Because $\frac{1}{c} \im\Sigma_c(\w)$ is nearly independent of $c$
for temperatures $T > T_\textrm{max}$ [cf.~Fig.~\ref{fig:Sc_c-U10}(b)],
also $c\tau(\w)$ does not depend on $c$ [see Fig.~\ref{fig:ftau-U10}(d)].

At temperatures $T < T_\textrm{max}$, conduction electrons and $f$
electrons tend to form heavy quasiparticles with Fermi-liquid properties,
leading to a small resistivity for $c=1$ and $T\to 0$. The introduction of
Kondo holes destroys the lattice periodicity and, therefore, the resistivity
increases; for $c\leq0.5$, the nonmonotonic course changes to a behavior which
is typical for a dilute distribution of impurity scattering centers
[cf.\ Fig.~\ref{fig:rho-t-U10}(a)]. As a consequence, the resistivity at
$T < T_\textrm{max}$ does not scale with the concentration $c$ of Kondo holes.
Instead, the normalized residual resistivity $\rho(T{=}0)/c$ depends
on $c$ and reaches a maximum at $c=0.5$ [cf.\ Fig.~\ref{fig:ftau-U10}(a)];
for small $c$ it can be approximated by
\begin{equation}
  \frac{r_0}{1-0.6c} \textrm{ with } r_0:=\lim_{c\to0}\frac{\rho(T{=}0)}{c}.
\label{eq:enhancement-factor}
\end{equation}
This is consistent with other calculations\cite{xu.li.1990,li.qiu.1991,mutou.2001b}
and can be understood using the CPA equation.
The approximation Eq.~(\ref{eq:relaxation-time-approx}) leads to
$\rho(T{=}0)/c= \sigma_0^{-1}/c\tau(0) \propto -\im \Sigma^\textrm{CPA}_c(0)/c$
[using the limit (\ref{eq:L_T0}) with abbreviation (\ref{eq:sigma_0})].
With the CPA equation (\ref{eq:cpa-efB_infty-Sigmaf}), we obtain
\begin{equation}
  \frac{1}{c} \im \Sigma^\textrm{CPA}_c(0)
    = \frac{1}{c} \im \frac{V^2}{-\Sigma^\textrm{CPA}_{ff}(0)}
    = \im \frac{V^2 G^A_{ff}(0)}{1 + c\Delta_A(0)G^A_{ff}(0)}.
\end{equation}
Because $\Delta_A(0)G^A_f(0)$ does not vanish, $\im \Sigma^\textrm{CPA}_c(0)/c$
as well as $\rho(T{=}0)/c$ acquire a $1/(1-ac)$ dependence.

A word is in order about the relation of the concentration dependence of the
resistivity to the typical scaling behavior for heavy-fermion materials,
e.g., in Ce$_x$La$_{1-x}$Cu$_6$.\cite{Onuki87} The reported magnetic
resistivity contribution normalized onto the Ce concentration decreases
monotonically with increasing cerium concentration.\cite{Onuki87}
In Ce$_x$La$_{1-x}$Cu$_{2.05}$Si$_2$, however, a nonmonotonic concentration
dependence of the magnetic part of the resistivity, normalized to its room
temperature value, was found.\cite{ocko.buschinger.etal.1999,ocko.drobac.etal.2001}
Our calculations describe consistently the crossover from a concentrated
lattice system to a dilute distribution of impurity scattering centers. As
discussed above, we can clearly pinpoint the origin of the maximum of
the normalized residual resistivity $\rho(0)/c$ at concentrations $c=x=0.5$.
It arises from multiple scattering processes at intermediate concentrations,
and the CPA equation yields a low-temperature enhancement factor of about
$1/(1-ac)$ [see Eq.~(\ref{eq:enhancement-factor})]. In the dilute limit,
$\rho(0)/c$ becomes independent of the concentration where the conduction
band self-energy is proportional to the single-particle
$t$-matrix.\cite{BickersCoxWilkins1987}

\begin{figure}
  \includegraphics[width=0.8\columnwidth]{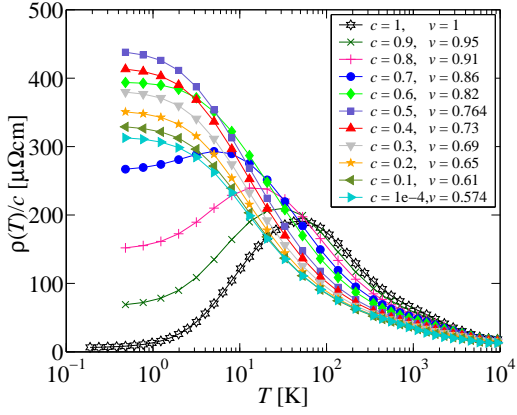}
  \caption{(Color online) Resistivity as in Fig.~\ref{fig:rho-t-U10}(b),
    but with $c$-dependent hybridization $V^2 = v(c) V_0^2$.}
\label{fig:rho-t-U10-VS}
\end{figure}

A weak dependence of the low-energy scale as a function of doping has been
reported in experiments.\cite{Onuki87,ocko.drobac.etal.2001} To simulate
these findings, we also studied an explicitly concentration-dependent
hybridization:
$V^2$ is scaled down in such a way that the same temperature scale $T_0$
is obtained for all concentrations. This is achieved by a linear
interpolation of $V^2$ between the concentrated limit and the dilute limit. Since
La increases the lattice spacing\cite{Onuki87,ocko.drobac.etal.2001}
of the materials, a decrease of the hybridization strength with increasing
concentration $1-c$ of the Kondo holes is consistent with the experimental
findings. In Fig.~\ref{fig:rho-t-U10-VS}, the resistivity is displayed for the
same parameters as in Fig.~\ref{fig:rho-t-U10}, but for $c$-dependent
hybridization. As a consequence, the calculated curves for the resistivity
scale with $cv(c)$. Since we do not know how to relate the explicit change
of the hybridization to a change of the unit cell volume, we are not able to
rescale the absolute unit.

The experimental determination of the resistivity faces some severe
limitations in determining the absolute value of magnetic contribution to the
resistance. Setting aside any problems stemming from the subtraction
of the phonon contributions by using a proper reference
material,\cite{Onuki87,ocko.buschinger.etal.1999} three major sources
of errors remain: (i) the accuracy of the geometry factor relating the measured
resistance to the resistivity, (ii) the accuracy of the determination of the
concentration $c$, and (iii) dealing with grain boundaries typical for
such heavy-fermion alloys. In particular, the error of the concentration value
$c$ increases significantly in the dilute limit, which has a profound impact
on the absolute value of $\rho(T)/c$. In addition grain boundaries in samples
might require the subtraction of the additional boundary resistivity in order
to extract $\rho_\textrm{mag}(T)$ instead of normalizing the resistivity to
its room temperature value, since grain boundaries are highly sample dependent.
Therefore, we limit ourselves to the reproduction of the qualitative features
of the reported experimental behavior.

\begin{figure}
  \includegraphics[width=0.8\columnwidth]{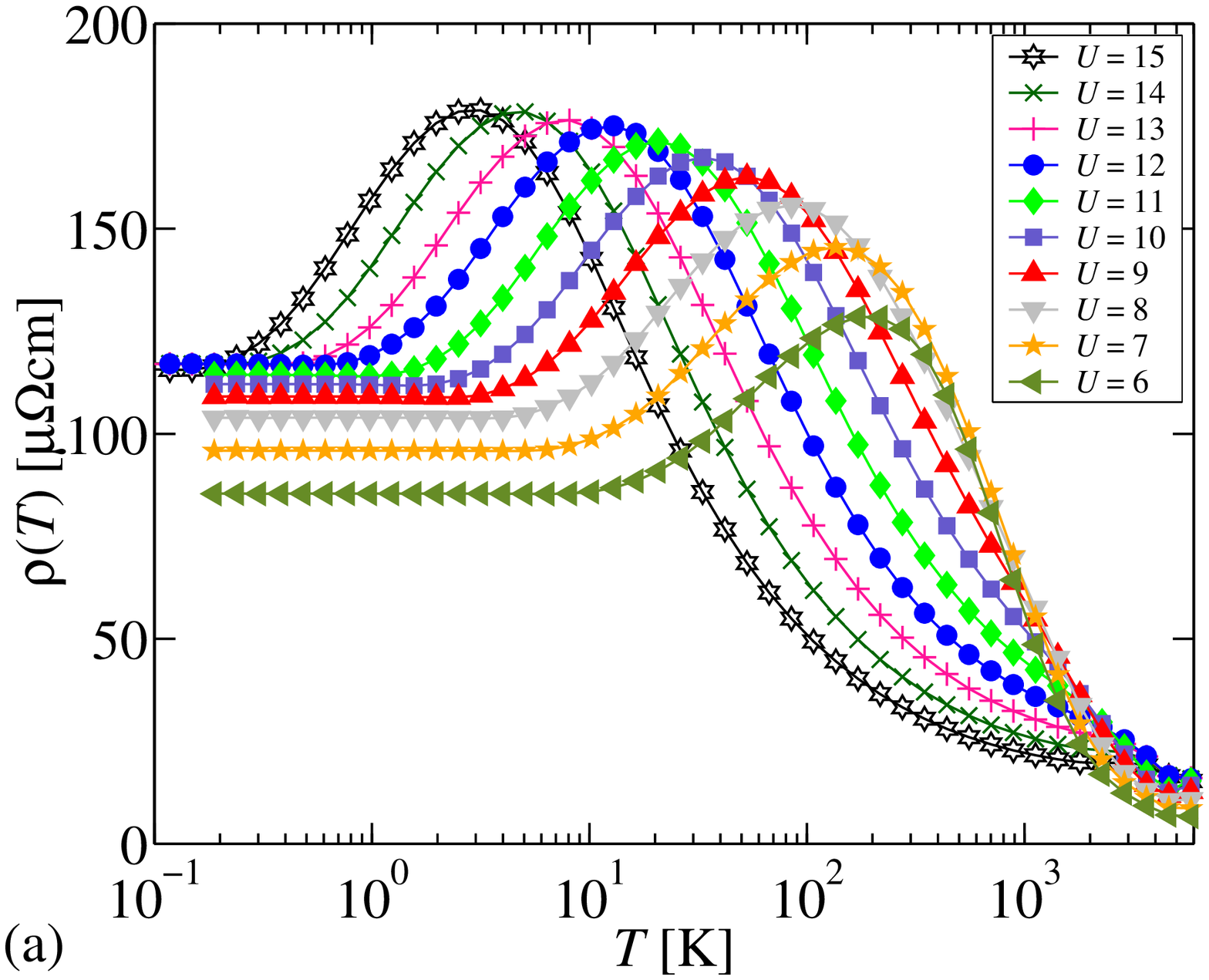}
  \includegraphics[width=0.8\columnwidth]{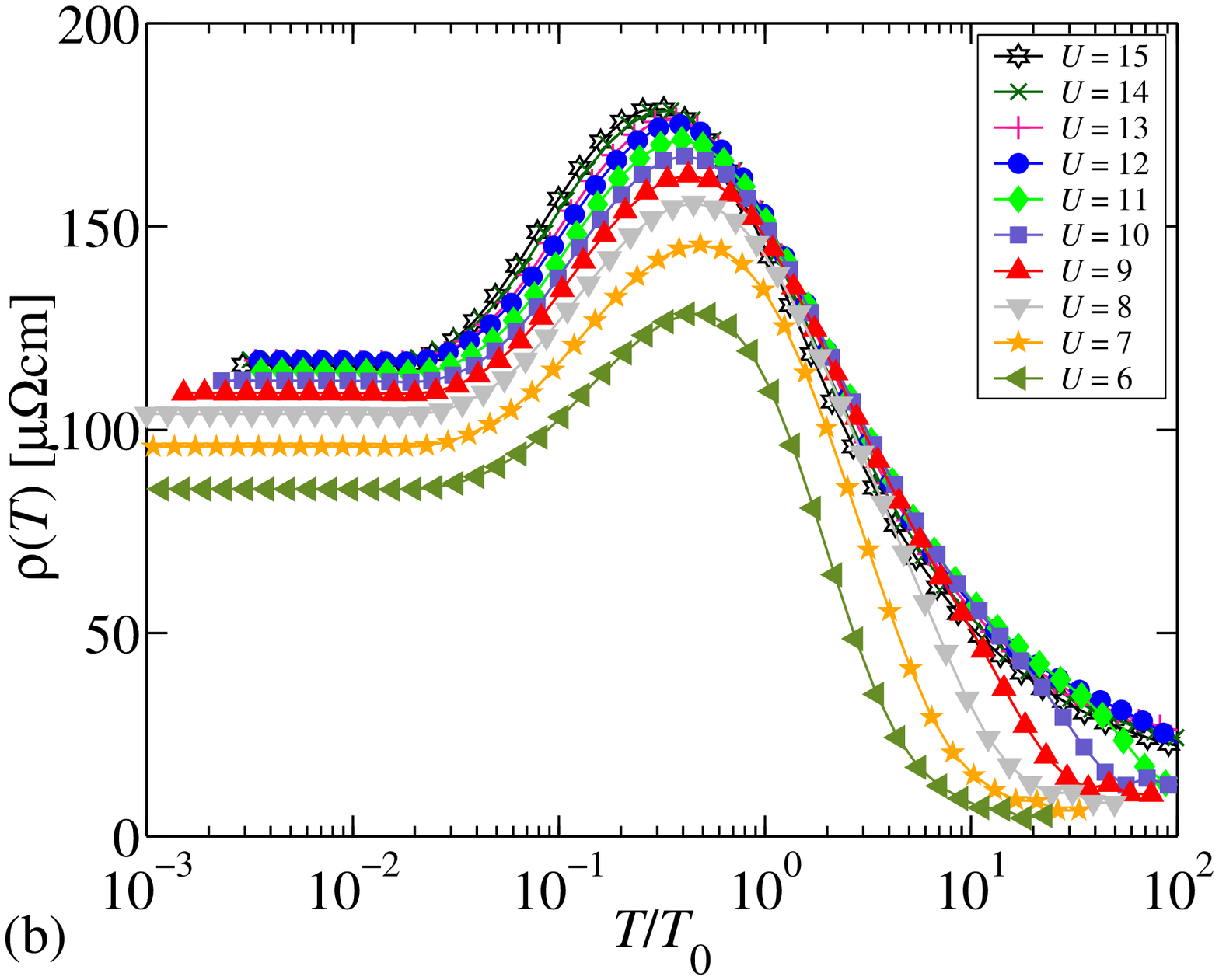}
  \caption{(Color online) Resistivity $\rho(T)$ as a function of $T$ for
    concentration $c=0.8$ of system $A$, calculated with CPA-NRG for different
    $U/\Gamma_0$, $\e^A_f-\e^A_c=-U/2$, $\e^B_f=\infty$,
    chemical potential $\mu=0$, and a filling $n_\textrm{tot}=1.4$.
    NRG parameters: number of retained NRG states, $N_s=800$,
    $\Lambda=1.6$, $\delta/\Gamma_0=10^{-10}$, and
    $L_w = 0$ (i.e., no Lorentzian broadening).}
  \label{fig:rho-t-c08}
\end{figure}

We also investigated the influence of the local Coulomb repulsion on
the transport properties. In Fig.~\ref{fig:rho-t-c08}, the resistivity
is displayed for a fixed concentration $c=0.8$, $\e^A_f-\e^A_c=-U/2$,
and various values of $U/\Gamma_0$.
The filling is set to $n_\textrm{tot}=1.4$ to match the previous setting
$n_\textrm{tot}=1.6-(1-c)$ in the case of fixed $U/\Gamma_0=10$.
Here, we reduced the small imaginary part $\delta/\Gamma_0$ to
$10^{-10}$ to reduce the error in the residual resistivity to a
minimum. While it turned out to be sufficient to use
$\delta/\Gamma_0=10^{-3}$ for values of the Coulomb interaction
up to $U/\Gamma_0=10$, this is no more the case for $U/\Gamma_0>10$.

For a particle-hole symmetric conduction band, no even-odd oscillations
in the NRG spectral function have been reported.\cite{peters2006} On the
contrary, we observe even-odd oscillations in the NRG spectral function
of the effective site with respect to the number of NRG iterations,
which is related to the temperature of the calculation in the usual
way.\cite{Wilson75,BullaCostiPruschke2007} This feature is related to
the two strong coupling fixed points\cite{Wilson75} of the NRG and
can show up for particle-hole asymmetric band density of states. We
used the standard technique for thermodynamical properties such as
entropy\cite{Wilson75,BullaCostiPruschke2007} and average the
resistivity as a function of temperature for even and odd numbers
of iterations.

The temperature $T_\textrm{max}$, at which the resistivity has its maximum,
is shifted to lower values for increasing $U$; one obtains an exponential
dependency of $T_\textrm{max}$ on $U$. For sufficiently strong $U$, the peak
height at $T_\textrm{max}$ is nearly independent of the value of $U$. This
is characteristic for the disordered stable moment regime, in which the
$f$ occupation reaches integer valence, and the increase of $U$ only reduces
the effective Kondo coupling $J\propto V^2/U$.
Charge fluctuations are strongly suppressed at low temperatures,
and one reaches a universal regime for a fixed concentration.
This is seen more clearly from Fig.~\ref{fig:rho-t-c08}(b), which
shows the scaling properties plotting the resistivity $\rho(T)$ versus
$T/T_0$. We have defined $T_0$ in Eq.~(\ref{eq:qp-weight-t0}) as an
effective low-temperature scale of subsystem $A$ in order to
clarify the scaling properties of transport properties. We certainly
do not imply universality of all low-energy properties.
We note, however, that $T_0$ is of the order of the position
$T_\textrm{max}$ of the maximum of the resistivity. While for
$U/\Gamma_0<8$, where high- and low-temperature scales are not very
well separated, the maximum of the resistivity and the peak height show
a $U$ dependence, and we reach a universality regime for large $U$.

\subsubsection{Thermoelectric power}

\begin{figure}
  \includegraphics[width=0.8\columnwidth]{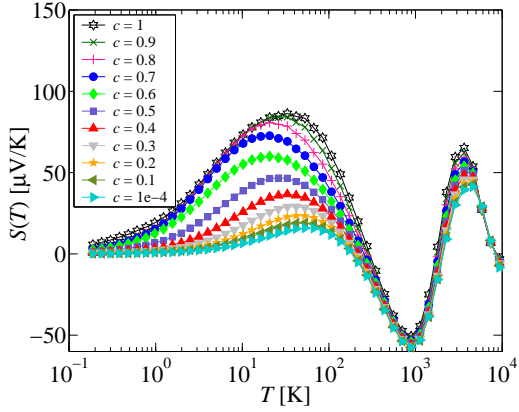}
  \caption{(Color online) Thermoelectric power $S(T)$ as a function of $T$
    for fixed $U/\Gamma_0=10$ and different concentrations $c$ of system $A$;
    all parameters as in Fig.~\ref{fig:rho-t-U10}, among others
    $n_\textrm{tot}=1.6-(1-c)$.}
\label{fig:s-t-U10}
\end{figure}

\begin{figure}
  \includegraphics[width=0.8\columnwidth]{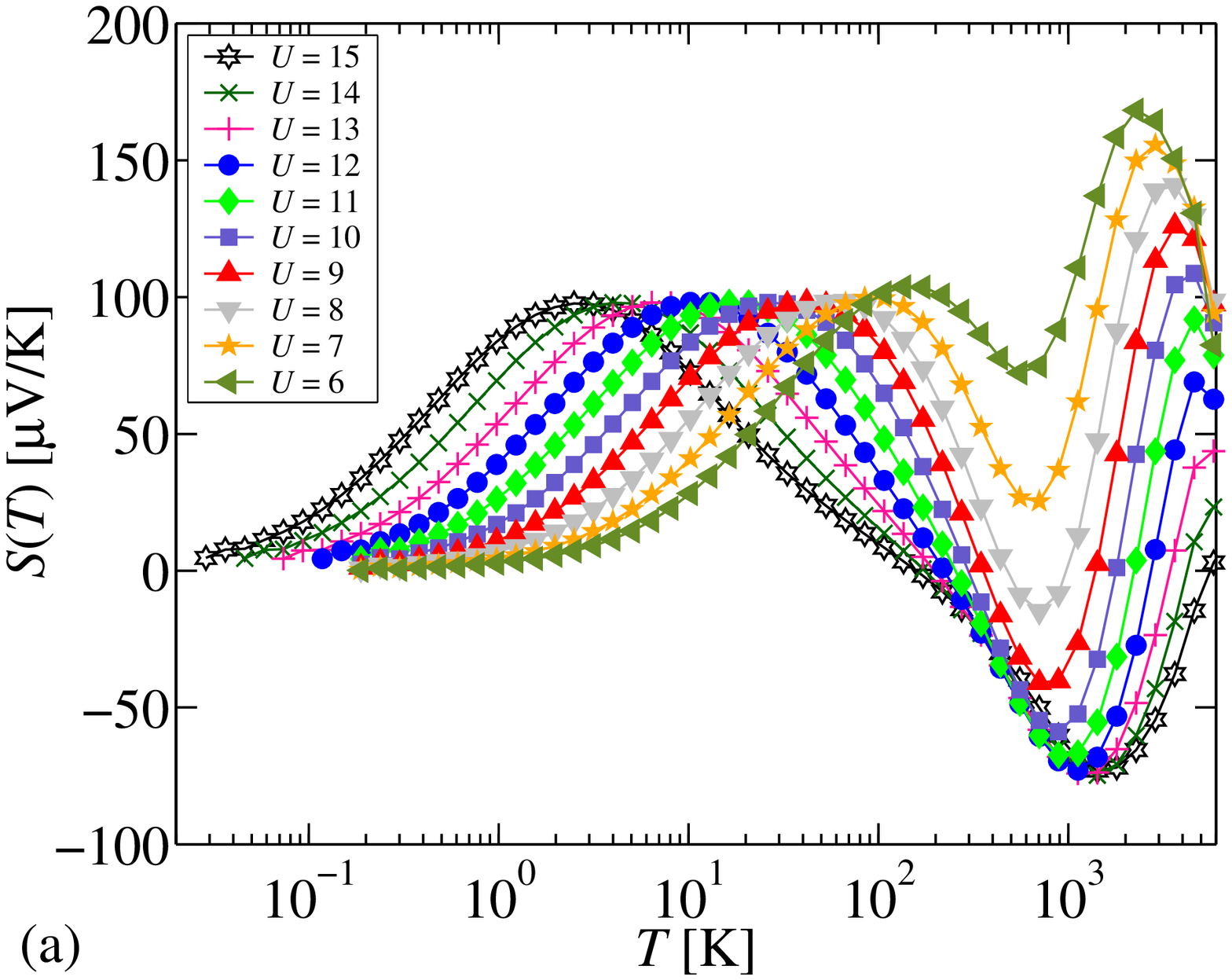}
  \includegraphics[width=0.8\columnwidth]{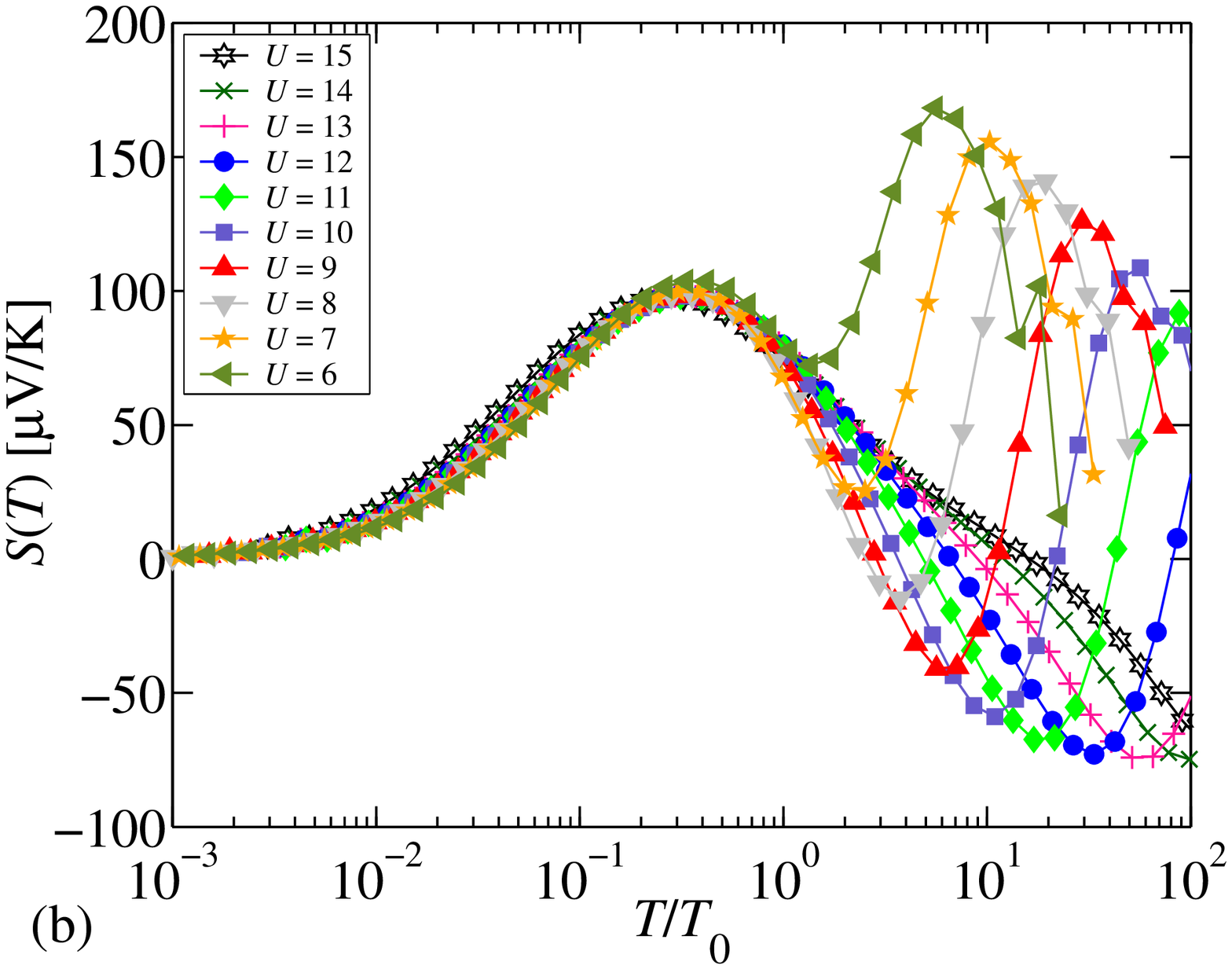}
  \caption{(Color online) Thermoelectric power $S(T)$ as a function of $T$
    for fixed concentration $c=0.8$ of system $A$ and different
    $U/\Gamma_0$; all parameters as in Fig.~\ref{fig:rho-t-c08},
    among others $n_\textrm{tot}=1.4$.}
\label{fig:s-t-c08}
\end{figure}

The thermoelectric power $S(T)$ measures the ratio between electrical and
heat current divided by the temperature, and its sign is related to the
integrated particle-hole asymmetry relative to the chemical potential.
In Fig.~\ref{fig:s-t-U10}, $S(T)$ is plotted for fixed $U/\Gamma_0=10$,
$\e^A_f-\e^A_c=-U/2$, and various values of the concentration $c$ of
system $A$.
The filling is kept at $n_\textrm{tot}=1.6-(1-c)$ and the $f$-level energy
of system $B$ is shifted to infinity.
This figure is accompanied by Fig.~\ref{fig:s-t-c08}.
Here, $S(T)$ is plotted for fixed filling $n_\textrm{tot}=1.4$, $c=0.8$,
$\e^A_f-\e^A_c=-U/2$, and various values of the Coulomb interaction $U$
[for the low-temperature scale $T_0$, see Eq.~(\ref{eq:qp-weight-t0})].

We obtain very large absolute values for $S(T)$ (of the magnitude
$100\;\mu\mathrm{V/K}$; see also Ref.~\onlinecite{pam.2006}). Note that the
thermoelectric power is obtained in absolute units, as already mentioned in
the beginning of this section. Similar to the resistivity, the thermoelectric
power exhibits a low-temperature peak which is correlated with the maximum of
the resistivity, which is an analytical consequence of Eq.~(\ref{eq:lin-res-thermo}).
Therefore, the position of this low-temperature peak depends on the
low-temperature scale $T_0$, which varies with the concentration $c$.
This can be seen in Fig.~\ref{fig:s-t-c08}(b), where the thermoelectric power
is shown on a rescaled axis $T/T_0$.

In addition, we observe a second extremum at a very high temperature
independent of the concentration and the low-temperature scale
which results from the charge fluctuations on the energy scale $\e_f-\mu$.
This maximum moves to higher temperatures with increasing $U$,
as shown in Fig.~\ref{fig:s-t-c08}.

\subsection{Disorder on the ligand sites}

In this section, we use the full matrix version of the CPA of
Sec.~\ref{sec:cpa}, where the two subsystems may have different parameter
matrices $\mathcal{V}_A$ and $\mathcal{V}_B$; i.e., we introduce
disorder on the ligand sites. We consider two cases: At first, we keep
the same hybridization for both subsystems, $V_A=V_B=V_0$; afterwards,
we use a site-dependent hybridization $V_A\neq V_B$.
In both cases, we fix the interaction at $U=10\Gamma_0$.

\subsubsection{Resistivity and thermopower for \texorpdfstring{$V_A=V_B=V_0$}{VA=VB=V0}}

\begin{figure}
  \includegraphics[width=0.8\columnwidth]{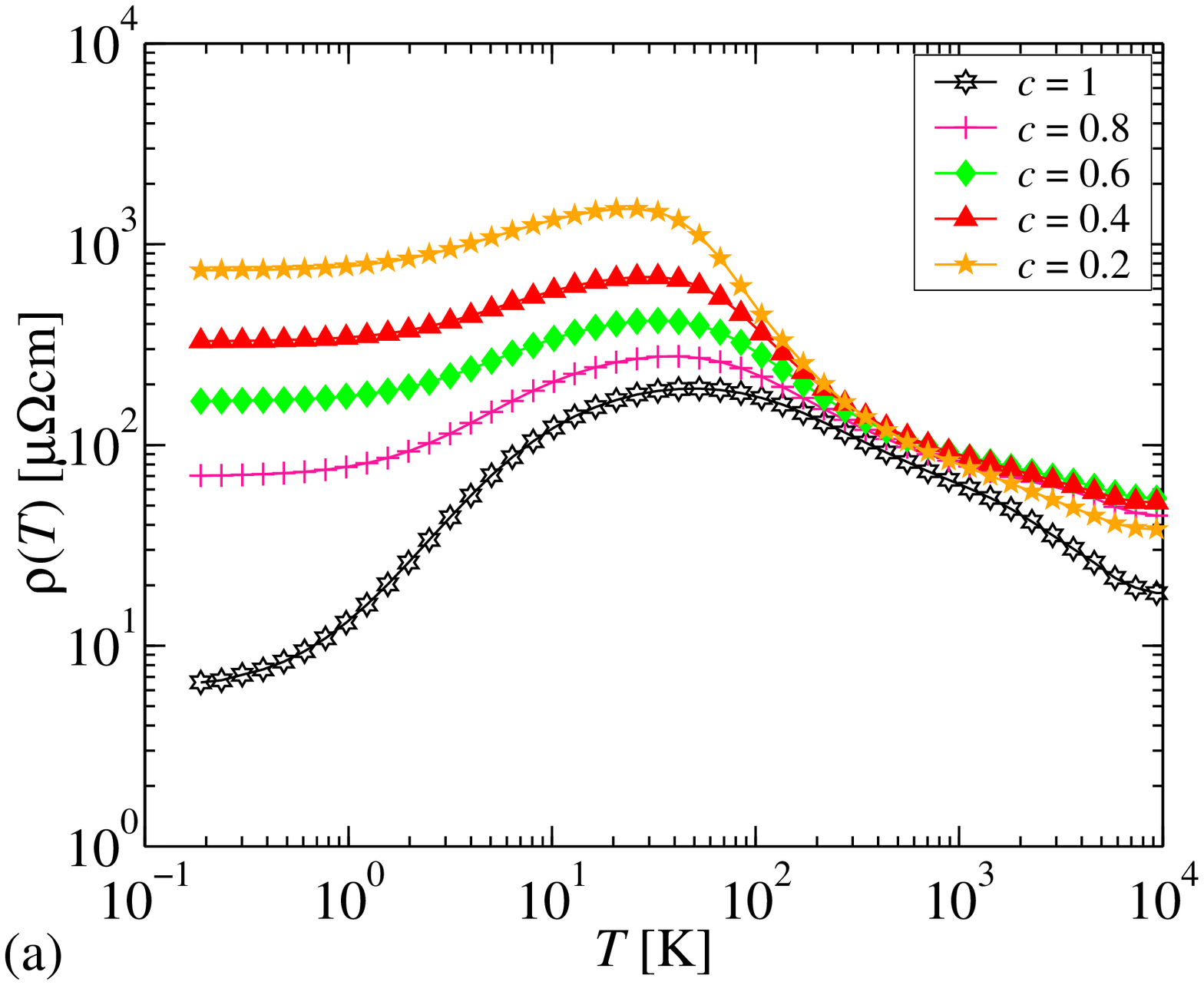}
  \includegraphics[width=0.8\columnwidth]{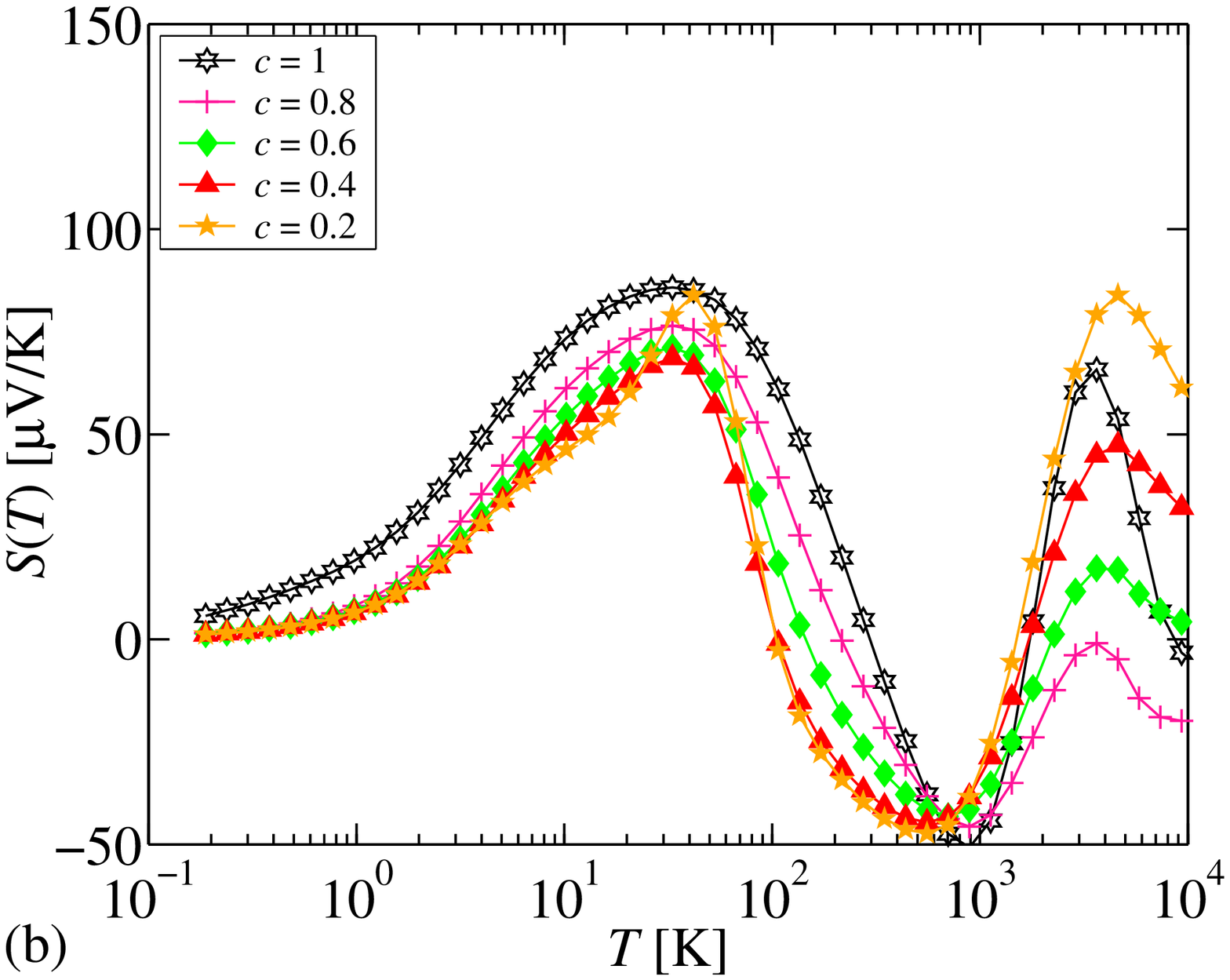}
  \caption{(Color online) (a) Resistivity $\rho(T)$ and (b) thermoelectric
    power $S(T)$ as functions of $T$ for different concentrations $c$ of
    system $A$, calculated with CPA-NRG for $V_A=V_B=V_0$, $U/\Gamma_0=10$,
    $\e^A_f-\e^A_c=-U/2$, $\e^B_f-\e^B_c=-\Gamma_0$, $\e^A_f=\e^B_f$,
    chemical potential $\mu=0$, and a filling $n_\textrm{tot}=2-0.4c$.
    NRG parameters: number of retained NRG states, $N_s=800$, $\Lambda=1.6$,
    and $\delta/\Gamma_0=10^{-3}$.}
  \label{fig:rhoS-t-band-n2}
\end{figure}

We start with the investigation of the transport properties as a function of
disorder in the case in which ligand substitution drives the system from
metallic to Kondo-insulating behavior. The concentration-dependent filling
is set to $n_\textrm{tot}=2-0.4c$, and the system evolves into a Kondo
insulator by reducing $c$. In Fig.~\ref{fig:rhoS-t-band-n2}, the resistivity
and the thermoelectric power are displayed for $\e^A_f-\e^A_c=-U/2$,
$\e^B_f-\e^B_c=-\Gamma_0$, and various values of the concentration $c$ of
system $A$. The curves for $c=1$ are identical to the $c=1$ curves in
Figs.~\ref{fig:rho-t-U10} and \ref{fig:s-t-U10}. We plot the resistivity on
a log-log scale in Fig.~\ref{fig:rhoS-t-band-n2}(a) to cover the almost 3
orders of magnitude of resistivity change.

In contrast to the previous case of $f$ disorder introduced by Kondo holes,
the number of Kondo scatterers remains constant as a function of the
concentration $c$. Therefore, the resistivity does not vanish for $c\to0$. For
temperatures $T>T_\textrm{max}$, the absolute values of the resistivity remain
nearly the same; above this characteristic low-temperature scale
$T_\textrm{max}$, lattice and Kondo disorder scattering become indistinguishable.
For temperatures $T<T_\textrm{max}$, an increase in the resistivity with
decreasing concentration $c$ is observed; Fig.~\ref{fig:rhoS-t-band-n2}(a)
shows a transition from low resistivity to an insulating behavior for $T\to0$
and $c\to0$. The disorder introduced in the conduction band destroys the
lattice coherence of the heavy quasiparticles at low temperatures.
Even though translational invariance is restored for $c\to0$, the residual
resistivity remains increasing for $c<0.5$ due to the crossover to an
insulating behavior.

The thermoelectric power does not show much variation with the concentration
$c$ as can be seen in Fig.~\ref{fig:rhoS-t-band-n2}(b). In contrast to the
resistivity, the thermoelectric power depends on the asymmetry of the spectrum
[see Eqs.~(\ref{eq:lin-res-thermo}) and (\ref{eq:L12})]. It turns out that the
asymmetry of the spectra as well as the asymmetry of the relaxation time as
defined in Eq.~(\ref{equ:relaxation-time}) change only very weakly with the
concentration, even though there is a significant renormalization of the
effective media $\Delta_{A/B}(z)$ [Eq.~(\ref{eq:medium-a-b})] which enters
the effective site calculations.

\begin{figure}
  \includegraphics[width=0.8\columnwidth]{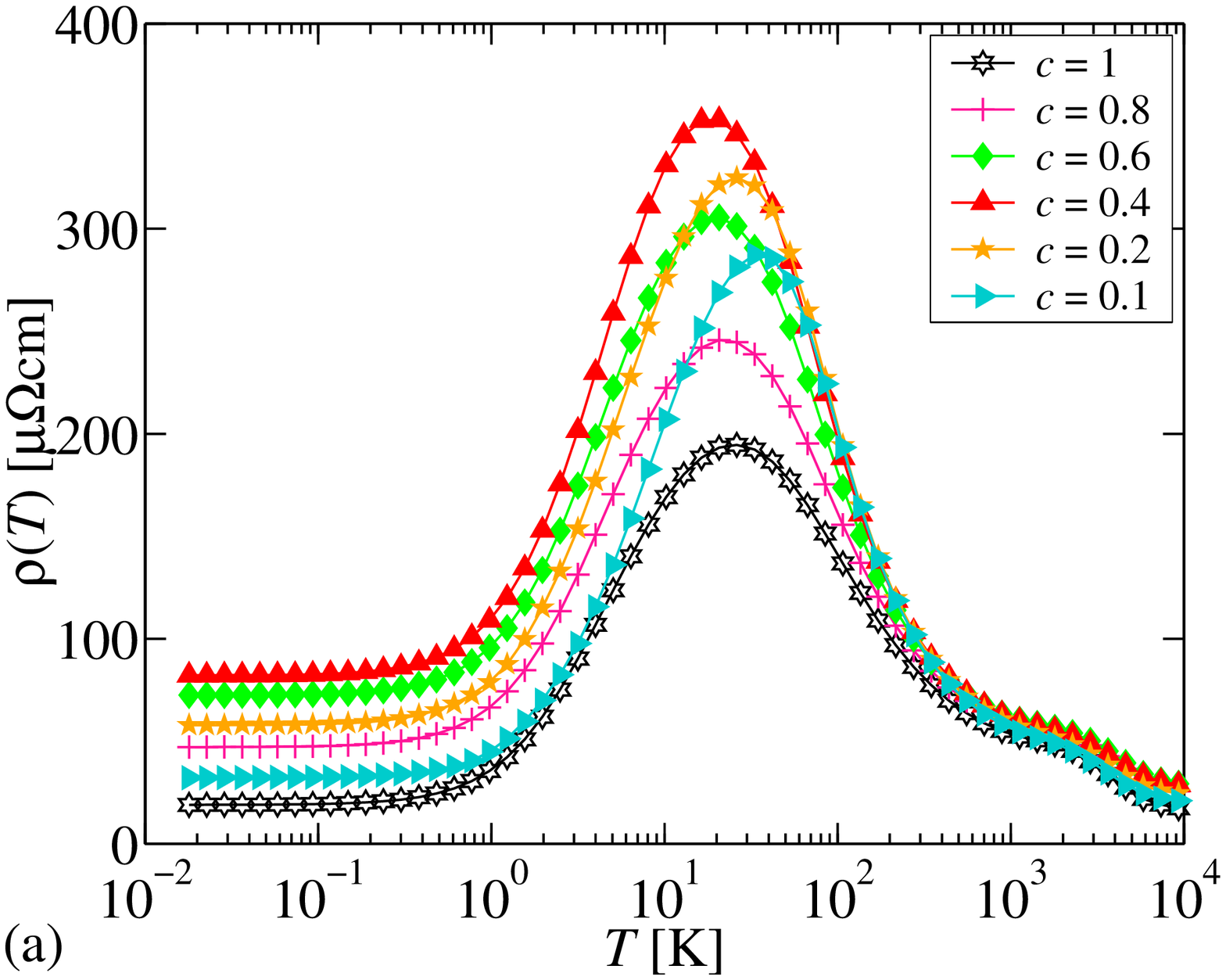}
  \includegraphics[width=0.8\columnwidth]{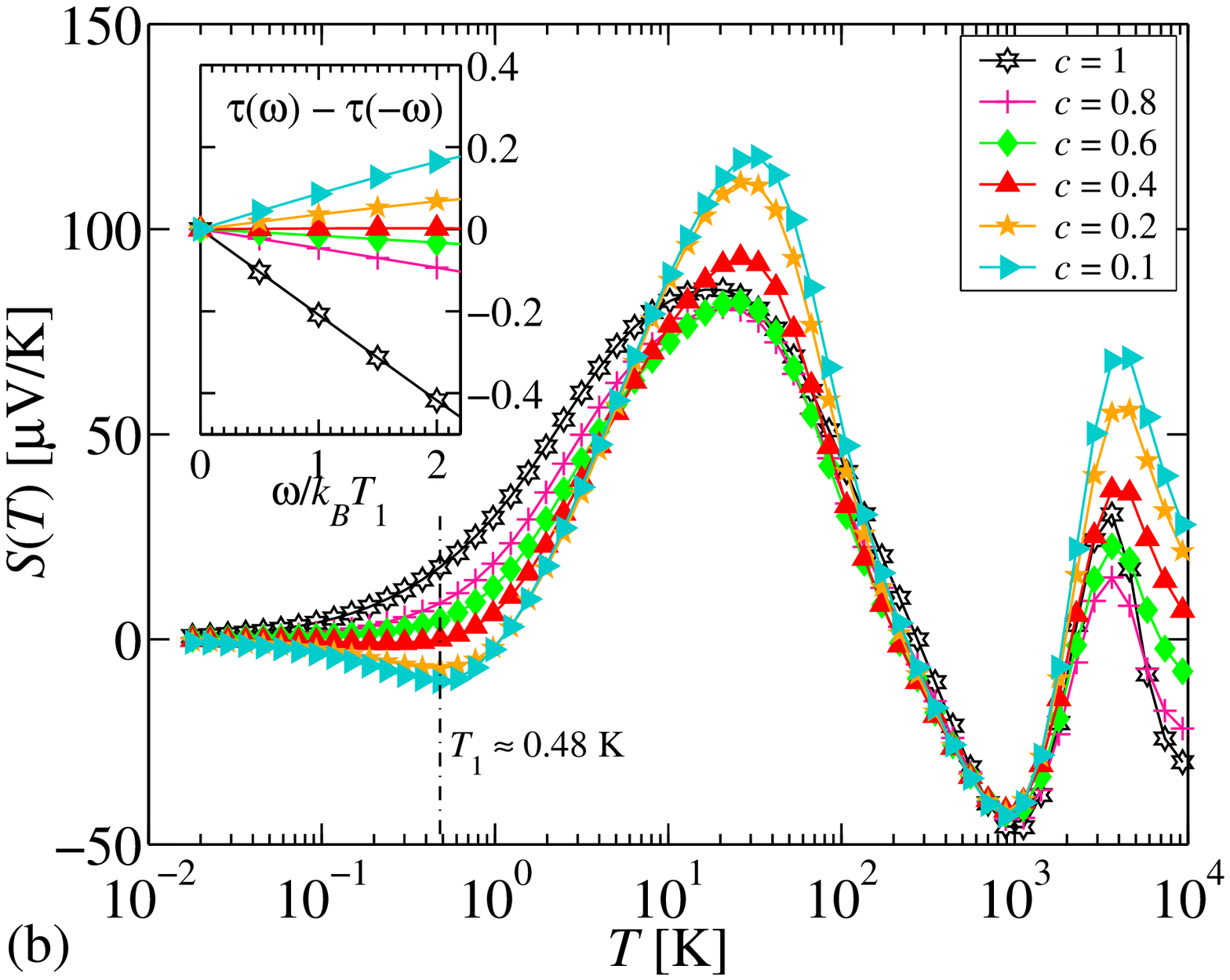}
  \caption{(Color online) (a) Resistivity $\rho(T)$ and (b) thermoelectric
    power $S(T)$ as functions of $T$ for different concentrations $c$ of
    system $A$, calculated with CPA-NRG for $V_A=V_B=V_0$, $U/\Gamma_0=10$,
    $\e^A_f-\e^A_c=-6\Gamma_0$, $\e^B_f-\e^B_c=-4\Gamma_0$,
    $\e^A_f-\e^B_f=0.2\Gamma_0$, chemical potential $\mu=0$, and a filling
    $n_\textrm{tot}=1.8-0.2c$.
    The inset of (b) shows the antisymmetrized relaxation time
    $\tau(\w)-\tau(-\w)$ [Eq.~(\ref{equ:relaxation-time})] in the vicinity
    of the chemical potential for the temperature $T_1=0.48\;\mathrm{K}$.
    NRG parameters: number of retained NRG states, $N_s=800$, $\Lambda=1.6$,
    and $\delta/\Gamma_0=10^{-3}$.}
\label{fig:rhoS-t-band}
\end{figure}

Next, we investigate the transport properties as a function of disorder in the
case in which ligand substitution keeps the system in a metallic regime for
all values of $c$. In Fig.~\ref{fig:rhoS-t-band}(a), the resistivity is
displayed for $\e^A_f-\e^A_c=-6\Gamma_0$ and $\e^B_f-\e^B_c=-4\Gamma_0$. In
this case, we have a disordered conduction band where both subsystems $A$ and
$B$ are asymmetric. By choosing a filling of $n_\textrm{tot}=1.8-0.2c$ and
$\e^A_f-\e^B_f=0.2\Gamma_0$, a nearly $c$-independent $f$ occupation
$n_f\approx0.92$ is achieved, significantly departed from the Kondo insulator
regime. This is clearly visible in the figure: in the limits $c=0$ and $c=1$,
the residual resistivity $\rho(T{\to}0)$ is small as expected for metallic
systems. For $0<c<1$, the disorder in the conduction band leads to an enhanced
resistivity. The residual resistivity peaks between $c=0.4$ and $0.6$. At high
temperatures, the resistivity becomes independent of the concentration $c$:
the incoherent Kondo scattering dominates the scattering processes over the
lattice disorder.

In Fig.~\ref{fig:rhoS-t-band}(b), the thermoelectric power is shown for the
same parameter set. The overall appearance is equal to the case with Kondo
holes, but, in addition, we observe a sign change at small temperatures. This
sign change originates from the change of asymmetry in the relaxation time
$\tau(\w)$ as defined in Eq.~(\ref{equ:relaxation-time}). To illustrate this
point, we plot the antisymmetrized relaxation time $F(\w):=\tau(\w)-\tau(-\w)$
in the vicinity of the chemical potential for $T_1=0.48\;\mathrm{K}$ as inset
in Fig.~\ref{fig:rhoS-t-band}(b). At this temperature, $S(T,c=0.1)$ reaches
its low-temperature minimum, while for $c=1$, a positive thermoelectric power is
found. For intermediate $c=0.4$, $F(\w)$ almost vanishes as can be seen in
the inset; simultaneously, the value of the thermoelectric power at $T_1$ is
very small. For concentrations $c<0.4$, the positive slope of $F(\w)$ leads
to a negative thermoelectric power, while for $c>0.4$, a negative slope is
found corresponding to a positive thermoelectric power. This change of
asymmetry is a consequence of the very subtle redistribution of spectral
weight also seen in the single particle spectra (not depicted here).

\subsubsection{Resistivity and thermopower for \texorpdfstring{$V_A\neq V_B$}{VA<>VB}}

In a real material, it is more likely that the hybridization changes with the
local environment. Two different scenarios could be discussed. The first one
considers each individual local configuration to be characterized by a
probability distribution function for the hybridization $P(V)$, which will
lead to a distribution of Kondo scales. This scenario was suggested as a
possible route to non-Fermi-liquid behavior in HFSs.\cite{PhysRevLett.69.1113}
We, however, restrict ourselves to $A$/$B$ ligand disorder, yielding one
effective hybridization for each type of lattice sites $A$ and $B$: $V_A$ and
$V_B$. For a homogeneous medium $\Gamma$ as defined in Eq.~(\ref{eq:gamma}),
this leads to two different energy scales $T_A$ and $T_B$ for the local
effective sites $A$ and $B$. We expect a crossover from one heavy fermion with
a lower characteristic energy scale to another with a higher value of $T_0$.

\begin{figure}
  \includegraphics[width=0.8\columnwidth]{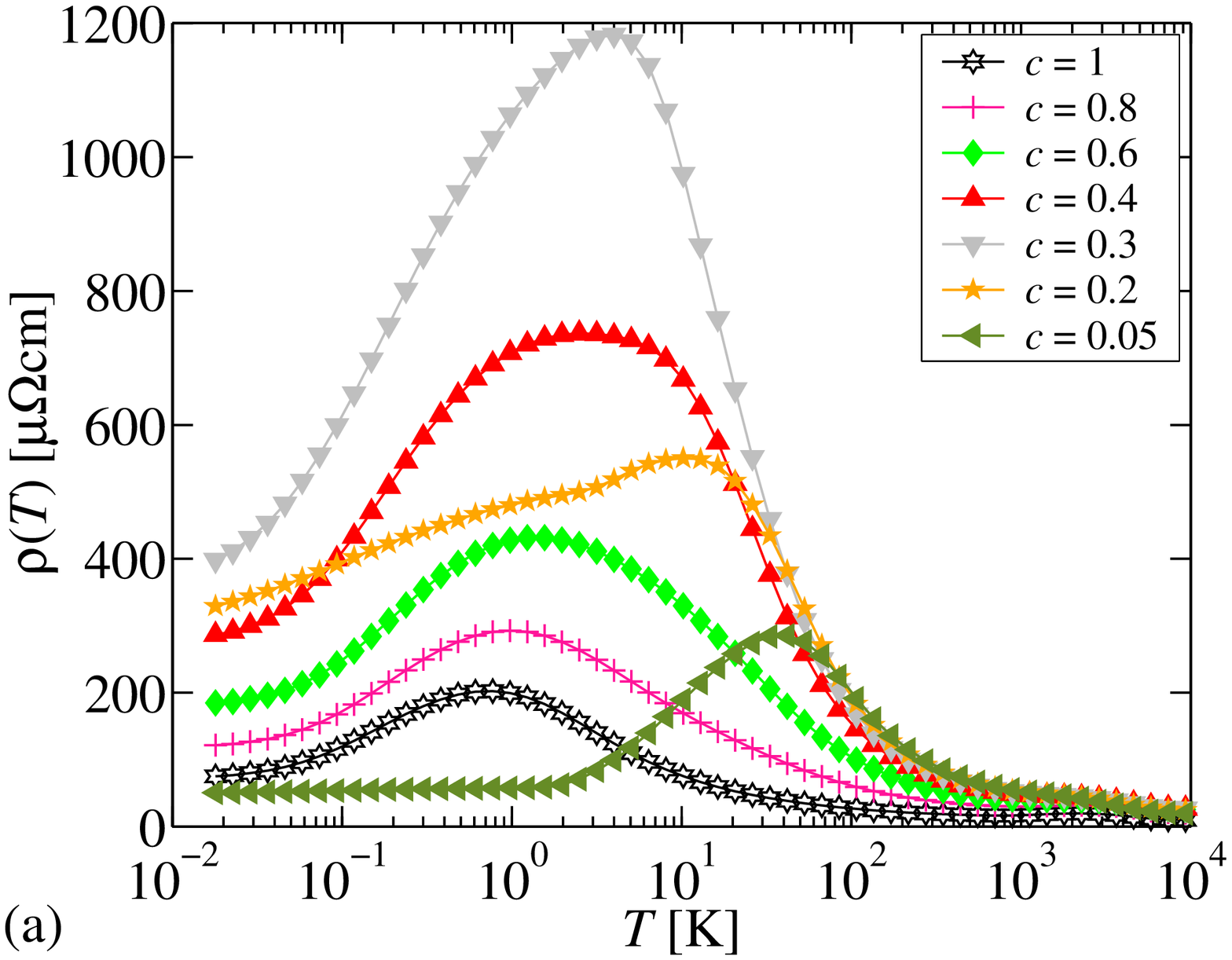}
  \includegraphics[width=0.8\columnwidth]{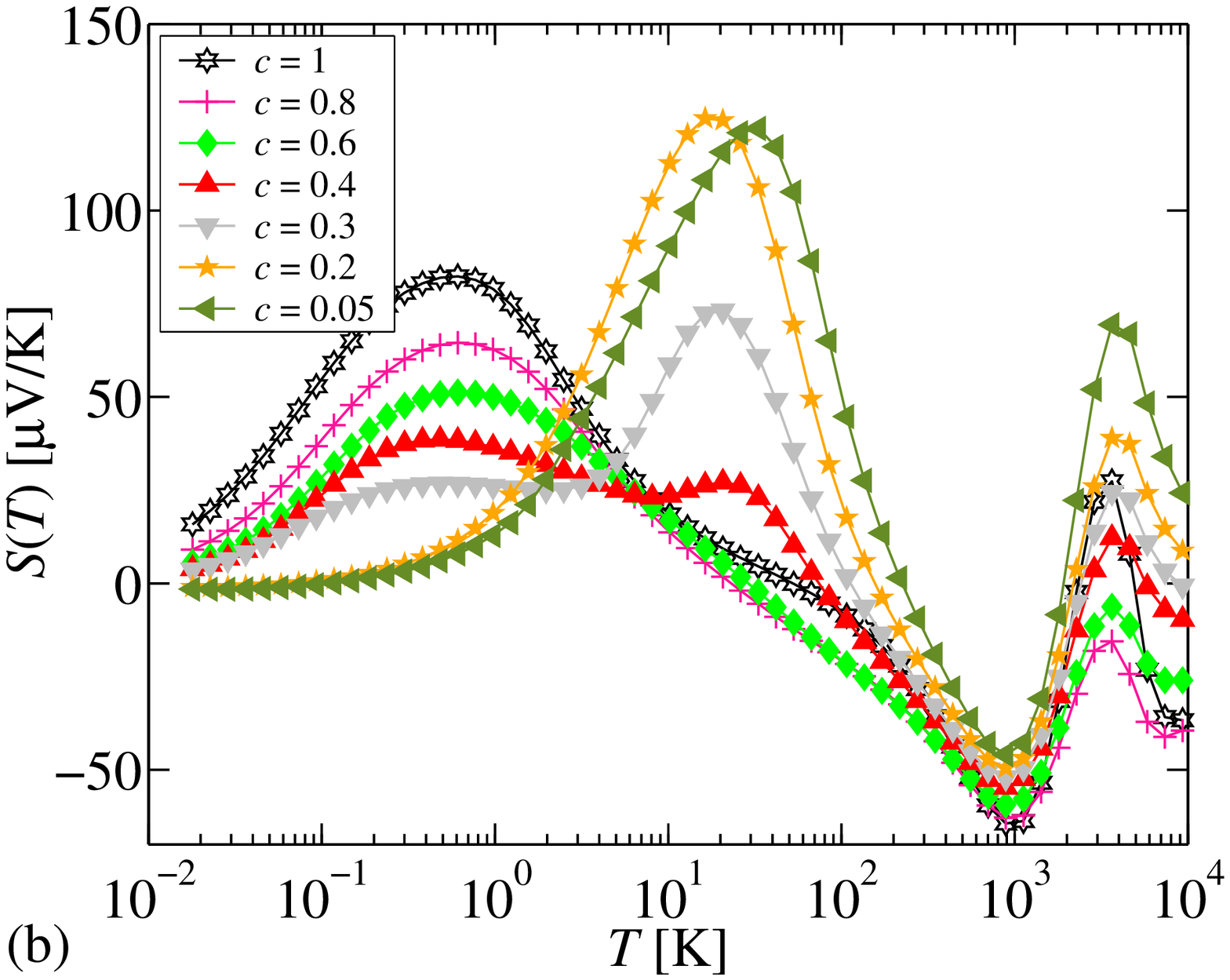}
  \caption{(Color online) (a) Resistivity $\rho(T)$ and (b) thermoelectric
    power $S(T)$ as functions of $T$ for different concentrations $c$ of
    system $A$; all parameters as in Fig.~\ref{fig:rhoS-t-band}, except for
    $V^2_A=3\Gamma_0^2$ and $\delta/\Gamma_0=10^{-4}$.}
\label{fig:rhoS-t-band-VA3}
\end{figure}

\begin{figure}
  \includegraphics[width=0.8\columnwidth]{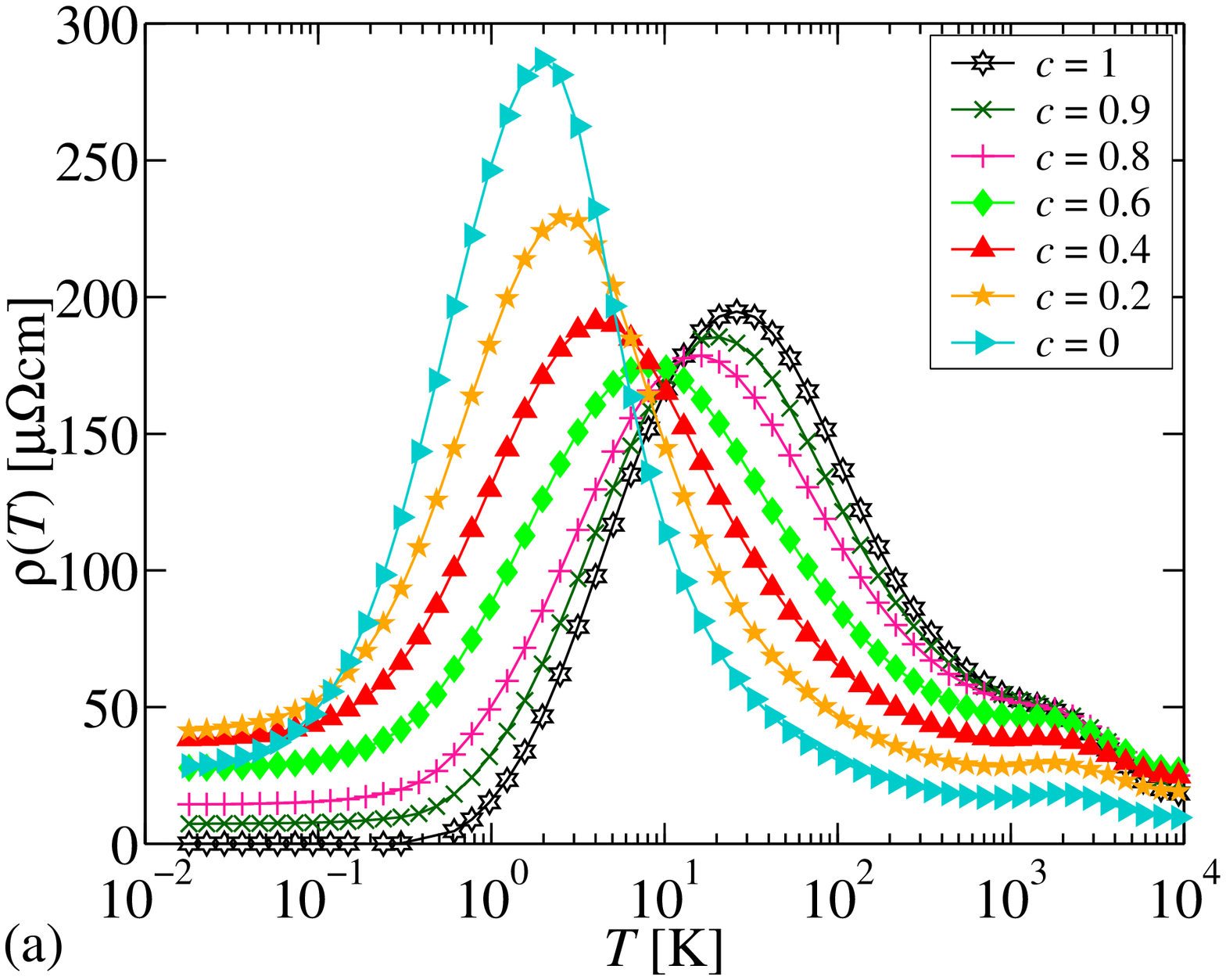}
  \includegraphics[width=0.8\columnwidth]{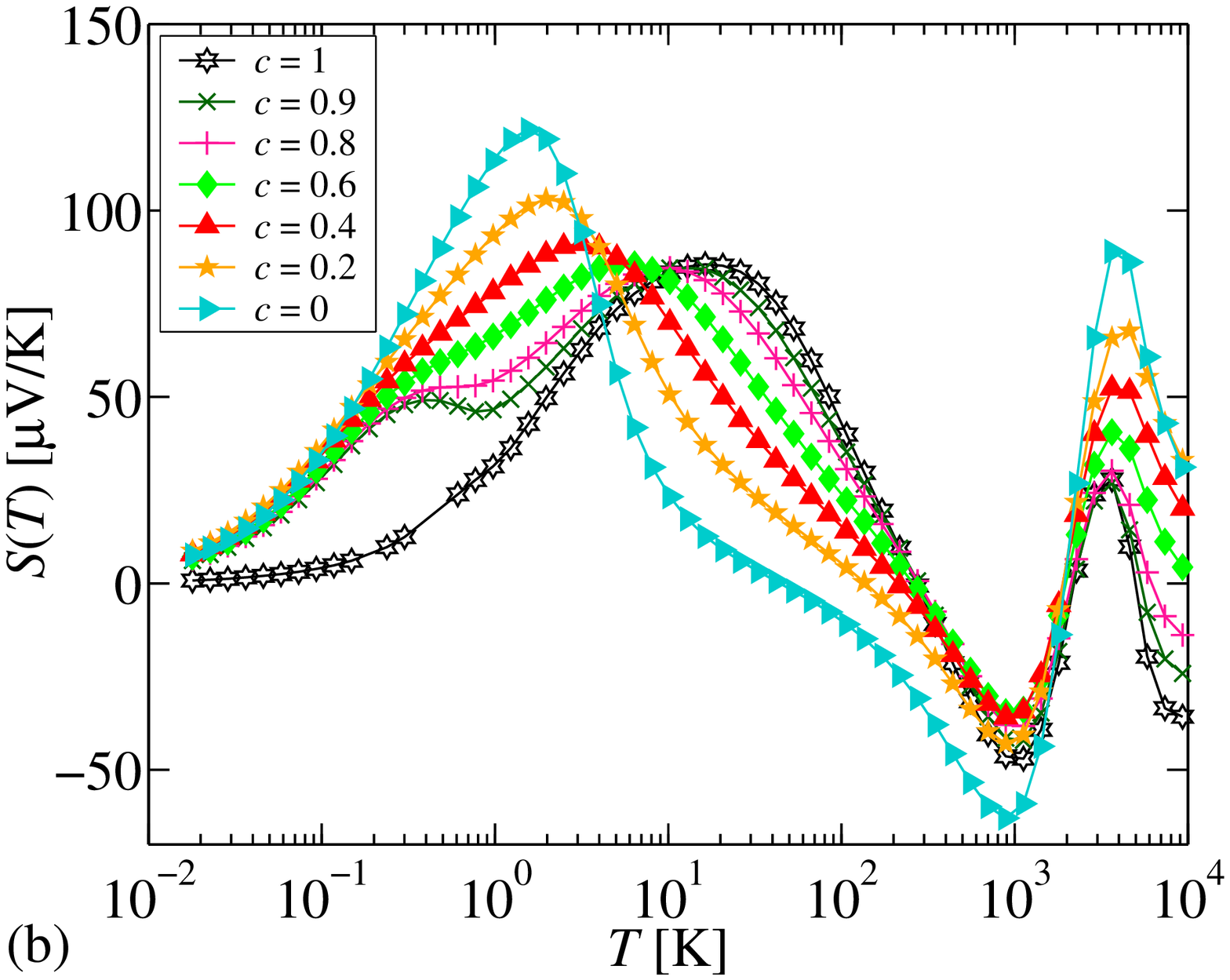}
  \caption{(Color online) (a) Resistivity $\rho(T)$ and (b) thermoelectric
    power $S(T)$ as functions of $T$ for different concentrations $c$ of
    system $A$; all parameters as in Fig.~\ref{fig:rhoS-t-band}, except for
    $V^2_B=3\Gamma_0^2$ and $\delta/\Gamma_0=10^{-4}$.}
\label{fig:rhoS-t-band-VB3}
\end{figure}

We take Fig.~\ref{fig:rhoS-t-band} as reference and adapt the choice of
parameters to it. There are two possibilities to achieve $V_A\neq V_B$:
we can change $V_A$ or $V_B$ with respect to Fig.~\ref{fig:rhoS-t-band}.
Both cases are discussed in the following. We reduced the small imaginary
part $\delta/\Gamma_0$ to $10^{-4}$ in order to reduce the error in the
residual resistivity.
In Fig.~\ref{fig:rhoS-t-band-VA3}, the resistivity and the thermoelectric
power for ligand disorder with changing hybridization from a low value of
$V^2_A=3\Gamma_0^2$ at $c=1$ to a high value of $V^2_B=V_0^2\approx5.64\Gamma_0^2$
are shown; in Fig.~\ref{fig:rhoS-t-band-VB3}, the hybridization conversely
changes from a high value of $V^2_A=V_0^2\approx5.64\Gamma_0^2$ at $c=1$ to a
low value of $V^2_B=3\Gamma_0^2$. All other parameters are the same as in
Fig.~\ref{fig:rhoS-t-band}.

In both figures, the CPA end points $c=0$ and $c=1$ nicely illustrate the
different energy scales, which are proportional to the temperature value of
the resistivity maximum and the corresponding maximum of the thermoelectric
power. With varying concentration, the thermoelectric power reflects
the crossover from one energy scale to the other by a decreasing height
of one maximum and an increasing height of the other one.
At intermediate concentrations, two maxima in $S(T)$ are found, which
clearly shows the presence of two energy scales. This feature is most
pronounced in Fig.~\ref{fig:rhoS-t-band-VA3} for $c=0.4$, where a shallow
double maximum structure is observed.

The behavior of the resistivity is completely different in the two cases
of Figs.~\ref{fig:rhoS-t-band-VA3} and \ref{fig:rhoS-t-band-VB3}. In
Fig.~\ref{fig:rhoS-t-band-VA3}, the residual resistivity increases with
decreasing $c$ and peaks around $c=0.3$. Due to the change of particle
numbers with the concentration, the residual resistivity is not symmetric
around $c=0.5$. At intermediate concentrations, we find an extended crossover
regime. At $c=0.2$, a non-Fermi-liquid behavior of the resistivity is observed
over more than three decades in temperature, while the thermoelectric power is
already dominated from the $B$ site transport.
Although the $c=0$ curves in Fig.~\ref{fig:rhoS-t-band-VB3} have the same
hybridization as the $c=1$ curves in Fig.~\ref{fig:rhoS-t-band-VA3}, they differ
in the number of electrons per unit cell, since in both cases $n_\textrm{tot}$
is given by $n_\textrm{tot}=1.8-0.2c$. The change in the resistivity as a
function of concentration appears to be much more gradually with a continuous
and monotonic shift of the position of the resistivity maximum.

\section{Conclusion and Outlook}
\label{sec:conclusion}

We have presented a detailed analysis of the transport properties of disordered
heavy-fermion compounds based on a combination of CPA to treat the disorder
and DMFT to handle the correlation effects. Our approach reduces to the
standard DMFT at those points corresponding to concentrated systems ($c=0$ and
$c=1$); i.e., our treatment of the transport properties interpolates between
two concentrated systems consisting only of pure $A$ or $B$ sites.

To make contact to typical experimental situations, we have studied two types
of local disorder: introduction of Kondo holes and disorder on the ligand
sites. In the case of doping with Kondo holes, we find an increasing
characteristic low-temperature scale $T_0$, defined via the quasiparticle
renormalization factor, as a function of the hole concentration for constant
hybridization. This is consistent with a previous comparison between the DMFT
and SIAM energy scales.\cite{PruschkeBullaJarrell2000,pam.2006} However,
in experiments, often only a weak dependence of the low-energy scale as a
function of doping has been observed.\cite{Onuki87,ocko.drobac.etal.2001}
Therefore, we investigated the change of the transport properties as a function
of concentration for a fixed effective low-energy scale. Such an ansatz is
consistent with an experimentally found increase of the lattice constants upon
Ce substitution by La, justifying a linear reduction of the hybridization
strength with increasing concentration of the Kondo holes.

The calculated resistivity normalized to the Ce concentration shows a
characteristic logarithmic upturn when lowering the temperature. In the
periodic system, the resistivity decreases again after a maximum is reached,
which is a sign of lattice coherence. Introducing Kondo holes reduces the
scattering of the $f$ electrons and gradually destroys lattice coherence. The
normalized resistivity continuously varies from that of a concentrated system
to that of a typical dilute Kondo scatterer. While we observe scaling with the
concentration at high temperature, multiscattering of conduction-band
electrons introduces an additional enhancement factor $1/(1-ac)$ in
the residual resistivity for $c\to 0$, which can become already important for
concentrations less than 10\%. Since the thermoelectric power is given
by a ratio of two transport integrals, it is much less sensitive to disorder.

The scaling properties of doped and undoped heavy-fermion materials appear to
be inconsistent with a simple Fermi-liquid theory and have caused a lot of
attention in recent years. The terminology ``non-Fermi liquid'' was coined for
those materials whose thermodynamic and transport properties are not well
understood---for a review, see Ref.~\onlinecite{Stewart01}. Usually, the
deviation from a $T^2$ law has been attributed to scattering of conduction
electrons on thermal and quantum fluctuations of the magnetic order parameter
in the vicinity of a quantum critical point. Such effects would generate a
$k$-dependent self-energy which is not included in our treatment. The effective
site for the Kondo-hole disorder problem reaches the strong-coupling fixed
point at low enough temperature in our CPA-NRG theory. Consequently, we know
analytically that the local self-energy $\Sigma^A(z)$ has Fermi-liquid
properties in this regime. The scaling regime for such a $T^2$ law, however,
turns out to be very small. The self-consistency condition and multiscattering
contributions to transport properties reorganize spectral weight in our
many-body calculation. Therefore, the crossover regime starts relatively early.
In this regime, one might be able to identify a rather extended temperature
range in which a power law with a fixed exponent $\alpha <2$ can be obtained
with the same accuracy as determined in experiments. However, the physical
meaning of such an exponent obtained in a crossover from a low-temperature
scaling regime is not apparent to us. Thus, we did not intend to extract such
a parameter.

In the case of ligand disorder, we have shown calculations which interpolate
between two metallic systems as well as describe the crossover from a Kondo
insulator to a metallic HFS. Each unit cell contains one Kondo scatterer which
experiences a different environment. A changing hybridization matrix element
leads to two different local low-temperature scales. For a constant
hybridization matrix element, we do not observe a shift in the position of the
resistivity maximum. The resistivity follows the typical heavy-fermion
behavior with a concentration-dependent residual resisitivity which peaks close
to the concentration of $c=0.5$. In the thermoelectric power, however, we note
a sign change at very low temperatures from positive values for $c<0.5$ to
negative values for $c=0.9$ similar to the one observed\cite{ocko.drobac.etal.2001}
in CeCu$_2$Si$_2$. The thermoelectric power is very sensitive to particle-hole
asymmetries in the spectral functions.

Our results clearly demonstrate that the combination of CPA and DMFT is capable
of accurately capturing the interplay of effects that are connected to local
correlations, on the one hand, and the average influence of disorder, on the
other hand. We discuss ligand disorder in addition to the conventionally
investigated introduction of Kondo holes. Such ligand disorder is relevant
for a variety of HFSs and, in particular, interesting in connection with
CeCu$_{6-x}$Au$_x$, the paradigm of non-Fermi-liquid behavior in the vicinity
of a quantum phase transition. Due to the lack of nonlocal fluctuations in the
DMFT-CPA treatment, it will not be possible to study the quantum critical
regime within our method. However, within our approach0 we can investigate the
influence of the ligand disorder on magnetic properties to obtain the magnetic
phase diagram. Another aspect is the role of orbital degrees of freedom, in
particular, on thermoelectric properties of HFSs. Again, the method introduced
here is valid for such an extended model, too, but the calculations become
much more involved. Investigations along these lines are in progress.

\begin{acknowledgments}
We thank R.~Bulla, J.~Freericks, N.~Grewe, A.~Hewson, U.~K{\"o}hler, D.~Logan,
N.~Oeschler, and V.~Zlati\'c for numerous discussions, and we thank the KITP
in Santa Barbara for its hospitality.
C.G., F.B.A., and G.C.\ acknowledge financial support by the Deutsche
Forschungsgemeinschaft, Project No.\ \mbox{AN 275/5-1}, and funding of the NIC,
Forschungszentrum J\"ulich, Project No.\ HHB000.
This research was also supported in part by the National Science Foundation
under Grant No.\ \mbox{PHY05-51164}.
\end{acknowledgments}


\end{document}